\documentclass[sn-basic]{sn-jnl}

\usepackage{graphicx}%
\usepackage{multirow}%
\usepackage{amsmath,amssymb,amsfonts}%
\usepackage{amsthm}%
\usepackage{mathrsfs}%
\usepackage[title]{appendix}%
\usepackage{xcolor}%
\usepackage{textcomp}%
\usepackage{manyfoot}%
\usepackage{booktabs}%
\usepackage{algorithm}%
\usepackage{algorithmicx}%
\usepackage{algpseudocode}%
\usepackage{listings}%

\usepackage{natbib}
\usepackage[normalem]{ulem}
\usepackage{color}

\theoremstyle{thmstyle}%
\theoremstyle{thmstyleone}%
\theoremstyle{thmstyletwo}%

\usepackage[permil]{overpic}
\graphicspath{{images/}}

\begin{document}

\title[Fully convolutional networks for velocity-field predictions based on the wall heat flux in turbulent boundary layers]{Fully convolutional networks for velocity-field predictions based on the wall heat flux in turbulent boundary layers}


\author*[1,2]{\fnm{Luca} \sur{Guastoni}}\email{guastoni@mech.kth.se}
\author[1,2]{\fnm{}Arivazhagan~G. \sur{Balasubramanian}}
\author[3]{\fnm{Firoozeh} \sur{Foroozan}}
\author[3]{\fnm{Alejandro} \sur{G\"uemes}}
\author[3]{\fnm{Andrea} \sur{Ianiro}}
\author[3]{\fnm{Stefano} \sur{Discetti}}
\author[1,2,4]{\fnm{Philipp} \sur{Schlatter}}
\author[5,2]{\fnm{Hossein} \sur{Azizpour}}
\author[1,2]{\fnm{Ricardo} \sur{Vinuesa}}

\affil[1]{\orgdiv{FLOW, Engineering Mechanics}, \orgname{KTH Royal Institute of Technology}, \orgaddress{\street{Osquars Backe 18}, \city{Stockholm}, \postcode{114 28}, \country{Sweden}}}

\affil[2]{\orgdiv{Swedish e-Science Research Centre (SeRC)}, \orgaddress{\street{Osquars Backe 18}, \city{Stockholm}, \postcode{114 28}, \country{Sweden}}}

\affil[3]{\orgdiv{Department of Aerospace Engineering}, \orgname{Universidad Carlos III de Madrid}, \orgaddress{\street{Avda. de la Universidad, 30}, \city{Legan\'es (Madrid)}, \postcode{28911}, \country{Spain}}}

\affil[4]{\orgdiv{Institute of Fluid Mechanics}, \orgname{Friedrich-Alexander Universit\"at}, \orgaddress{\street{Cauerstraße 4}, \city{Erlangen-N\"urnberg}, \postcode{91058}, \country{Germany}}}

\affil[5]{\orgdiv{Division of Robotics, Perception, and Learning}, \orgname{KTH Royal Institute of Technology}, \orgaddress{\street{Lindstedtsv\"agen 30}, \city{Stockholm}, \postcode{114 28}, \country{Sweden}}}


\abstract{Fully-convolutional neural networks (FCN) were proven to be effective for predicting the instantaneous state of a fully-developed turbulent flow at different wall-normal locations using quantities measured at the wall. In Guastoni \textit{et al.} (\textit{J. Fluid
Mech.} 2021, 928, A27), we focused on wall-shear-stress distributions as input, which are difficult to measure in experiments. In order to overcome this limitation, we introduce a model that can take as input the heat-flux field at the wall from a passive scalar. Four different Prandtl numbers $Pr = \nu/\alpha = (1,2,4,6)$ are considered (where $\nu$ is the kinematic viscosity and $\alpha$ is the thermal diffusivity of the scalar quantity). A turbulent boundary layer is simulated since accurate heat-flux measurements can be performed in experimental settings: first we train the network on aptly-modified DNS data and then we fine-tune it on the experimental data. Finally, we test our network on experimental data sampled in a water tunnel. These predictions represent the first application of transfer learning on experimental data of neural networks trained on simulations. This paves the way for the implementation of a non-intrusive sensing approach for the flow in practical applications.}




\maketitle

\section{Introduction}
\label{sec:intro}
Several approaches have been proposed over the years in order to perform a reliable non-local estimation of the velocity field in wall-bounded flows. When considering linear methods, an optimal solution can be obtained using extended proper orthogonal decomposition~\cite[EPOD, ][]{boree2003extended}, which is equivalent to linear stochastic estimation~\cite[LSE, ][]{miguel}. Recent works on transfer functions~\citep{sasaki_2019} have highlighted the benefits of using non-linear methods. In this regard, neural-network models have shown excellent results in monitoring the instantaneous state of the flow using quantities measured at the wall since the seminal work of \cite{milano2002neural} and as recently highlighted in our contributions \cite[see, e.g.][]{guemes2019sensing, guastoni2020prediction, guastoni_2021, guemes_2021}. 
Fully-convolutional networks (FCNs) provide an accurate reconstruction of the flow field at a given wall-normal location, when highly-resolved wall-shear-stress fields are used as inputs~\citep{guastoni_2021}. Other architectures have been tested in the literature for the same task, \textit{e.g.} super-resolution generative adversarial networks~\citep[SR-GANs, ][]{guemes_2021}. 
While it is straightforward to measure the wall-shear-stress components in direct numerical simulations (DNSs), sampling the same quantities in an experiment is much more difficult, in particular if a high resolution is required. This highlights the need of other input quantities, whose acquisition is more practical in experiments. 
In particular,~\cite{ABE2004404} report significant similarities between the wall-normal heat flux fluctuations and the streamwise wall-shear-stress fluctuations.
Several examples in the literature demonstrate the feasibility of time-resolved convective heat flux measurements~\citep{hetsroni1994heat,gurka, nakamura, raiola2017towards}, hence this quantity is used by our neural-network model to reconstruct the flow field.
Previously,~\cite{kim_lee_2020} used convolutional neural networks (CNNs) to predict the instantaneous wall-normal heat flux from the two wall-shear-stress components. Their approach still relies on the knowledge of the shear stresses at the wall and both the measurements and the predictions are at the same location.
In this work, on the other hand, we consider input measurements at the wall and target flow-state estimation above the wall.

In this work we aim to use FCNs to reconstruct the experimental data sampled in the water-tunnel facility at Universidad Carlos III de Madrid using InfraRed (IR) thermography for convective heat transfer \citep{astarita2012infrared} and Particle Image Velocimetry \cite[PIV,][]{raffel2018particle} for velocity-field measurements.
Despite machine-learning-based control having been tested in experimental settings with promising results~\citep{gautier_2015}, it is difficult to acquire from experimental facilities the large datasets that are needed to train neural network models. Additionally, experimental uncertainty needs to be taken into account and the possibility to assess the effect of the spatial resolution of the samples is limited. For these reasons, in this study, we perform the training of the networks using the data obtained from numerical flow simulations. A zero-pressure-gradient turbulent boundary layer is simulated, matching the experimental values in terms of Prandtl and Reynolds numbers.
First, we consider flow fields sampled from DNSs at full resolution, performing predictions of increasing difficulty depending on the inputs of the neural network. Second, we cut, filter and downsample the DNS data to match the characteristics of the experimental data. The neural-network models are optimized using these synthetic experimental data. Finally, the trained neural networks are tested on the data from the water tunnel.

After this introduction, the paper is organized as follows. In section~\ref{sec:dataset}, the setup of the numerical simulation of a zero-pressure-gradient turbulent boundary layer is described, along with the number of fields that are sampled for training, validation and testing. The experimental setup is also described in this section. In section~\ref{sec:meth}, the neural-network architecture and training details are reported, as well as the preparation of DNS data to mimick the experimental ones. In section~\ref{sec:results}, the performance of several neural networks with varying numbers of convolutional layers is compared for predictions with different inputs. The capability of the neural networks to reconstruct the synthetic and real experimental data is also analyzed. Finally, concluding remarks and future research directions can be found in section~\ref{sec:conc}.

\section{Dataset}
\label{sec:dataset}
\subsection{Numerical dataset}
The direct numerical simulation (DNS) from which the measurements and the target output fields are sampled is performed using the pseudo-spectral code SIMSON~\citep{simson}. While our previous work~\citep{guastoni_2021} focused on a fully-developed flow, namely a turbulent open-channel flow, in this work we simulate a zero-pressure-gradient (ZPG) turbulent boundary layer (TBL).  
The inflow condition for the velocity is a laminar profile. A random trip forcing is applied to trigger the transition to a turbulent boundary layer. A fringe forcing is applied at the outflow in order to achieve periodicity at the boundary, as requested by the solution method. Four passive scalars representing the temperature of the fluid are also simulated. We consider Prandtl number $Pr=(1,2,4,6)$, indicating them with $(\theta_1,\theta_2,\theta_3,\theta_4)$, respectively. For all the passive scalar we impose an isothermal wall boundary condition $\theta_i |_{y=0} = 0$, for $i=1,2,3,4$.
The highest Prandtl number $Pr=6$ is the result of a trade-off between the value that can be measured in our experimental setting and the computational cost of simulating such a flow with a DNS. The higher the Prandtl number, the smaller the thermal boundary layer, increasing the simulation resolution required to resolve all the relevant turbulent scales. The use of the same model for different Prandtl numbers allows us to investigate how the different thermal diffusivity influences the reconstruction performance.
The choice of a spatially-developing flow implies an additional degree of complexity in the predictions with respect to the previously-studied channel flow, since the friction Reynolds number $Re_{\tau}$ (based on the boundary-layer thickness and the friction velocity $u_{\tau}=\sqrt{\tau_w/\rho}$, where $\tau_w$ is the wall-shear stress and $\rho$ is the fluid density) increases with the streamwise location $x$ within the sampled fields. The highest considered $Re_{\tau}$ is 396, which is similar to the $Re_\tau$ of our experimental setting.
Few numerical simulation results are reported in the literature with a similar highest Prandtl number $Pr$ and friction Reynolds number $Re_{\tau}$. \cite{alcantra_0p71} simulated a channel flow resolving all the turbulent scales with $Pr=6$ and $Re_{\tau}=500$. \cite{kozuka} performed a DNS of a turbulent channel flow at lower Reynolds number $Re_{\tau}=395$, however, the maximum considered Prandtl number was $Pr=7$. A statistical characterization of the scalars and the comparison with the results from the previously-cited works are available in the work by~\cite{Balasubramanian_Guastoni_Schlatter_Vinuesa_2023}.

In our simulations, we sample the wall-shear-stress components, as well as the wall pressure.  Note that we considered a reference friction velocity at the middle of the computational domain, which implies that the actual inner-scaled location that is actually sampled slightly varies along the streamwise direction. However, the variation is within $\pm 0.1y^{+}$. Here the '+' denotes viscous scaling, {\it i.e.} in terms of the friction velocity $u_{\tau}$ or the viscous length $\ell^{*}=\nu / u_{\tau}$ (where $\nu$ is the fluid kinematic viscosity).
Furthermore, the flux of a passive scalars $\partial\theta_i/\partial y$ is sampled at the wall. 
The velocity-fluctuation fields (whose streamwise, wall-normal and spanwise components are denoted as $u,~v$ and $w$, respectively) are sampled at four wall-normal locations: $y^{+}=15,30,50$ and $100$.

\begin{figure}
    \begin{center}
    \includegraphics[width=.9\textwidth]{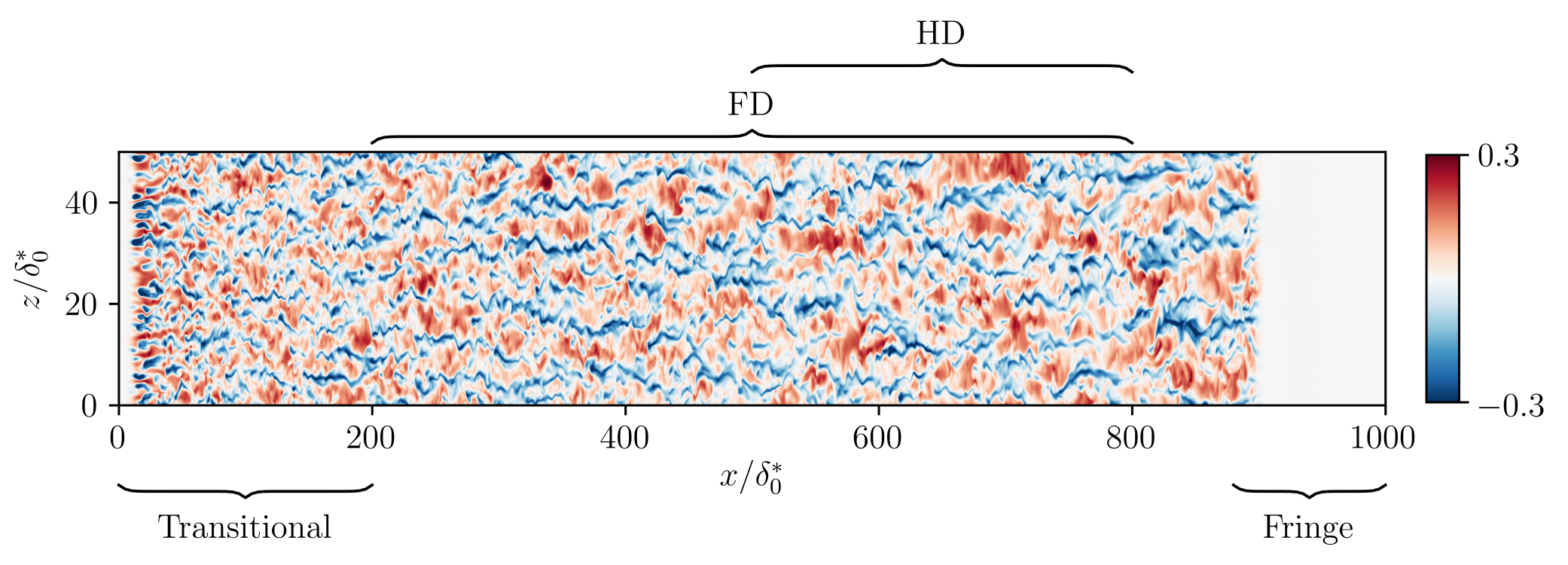}
    \end{center}
    \caption{\label{fig:fd_and_hd} Representation of full domain (FD) and half domain (HD) in a sampled streamwise velocity-fluctuation field at $y^{+}=30$.}
\end{figure}

Note that the sampled fields also include both the initial, transitional part of the flow and the final region affected by the fringe forcing. On the other hand, the neural-network models predict only a portion of the field. Depending on the size, we can identify two different types of samples, as shown in figure~\ref{fig:fd_and_hd}: \textit{full domain} (FD) samples have streamwise and spanwise lengths of $x_s/\delta^*_0=600$ and $z_s/\delta^*_0=50$, respectively. Here $\delta^*_0$ is the displacement thickness of the laminar boundary layer at the inflow, defined as:
\begin{equation}
    \delta^*_0 = \int_0^{\infty} \left(1 - \frac{u(0,y)}{U}\right)dy,
\end{equation}
with $U$ indicating the free-stream velocity. The samples do not include the initial ($x/\delta^*_0<200$) and the final region ($800<x/\delta^*_0<1000$). When the streamwise length of the samples is reduced to $x_s/\delta^*_0=300$, we refer to them as \textit{half-domain} (HD) samples.
The grid points considered in the FD case are $N_{\rm x,s} \times N_{\rm z,s} = 1960 \times 320$, while for HD they are $980 \times 320$.

In order to obtain a sufficiently large number of fields, five different realizations of the simulation are performed using five trip-forcing at the same location, but with different random seeds. The \textit{training dataset} includes 7,474 samples obtained from three of the five simulations. The \textit{validation dataset} consists of 2,195 samples, taken from a separate DNS to avoid unwanted correlations with the training dataset. The \textit{test dataset} is obtained from the remaining simulation, it includes 1,973 samples and the overall sampled time for testing is sufficient to obtain converged turbulence statistics.

\subsection{Experimental dataset}
\label{ssec:exp_data}
\begin{figure}
    \begin{center}
    \includegraphics[width=.45\textwidth]{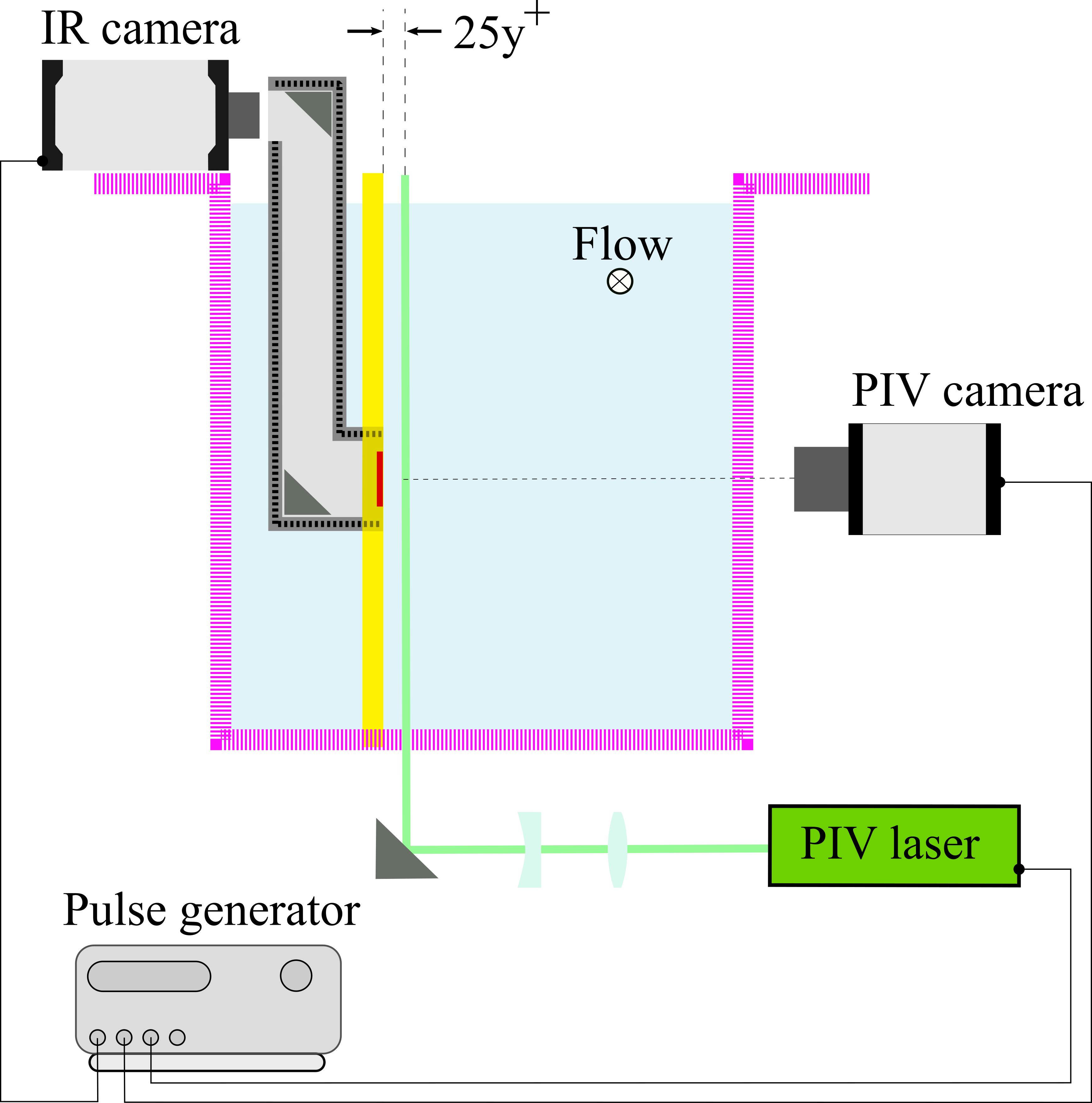}
    \end{center}
    \caption{Sketch of the experimental setup in lateral view (adapted from \citealt{foroozan}). The flat plate is indicated in yellow, the water-tunnel walls in magenta, and the periscope box in grey. PIV and IR measurement planes are shown in green and red, respectively.}
    \label{fig:setup} 
\end{figure}

This section describes the experimental setup designed to obtain a ZPG turbulent boundary layer, as well as the measurement techniques employed to provide simultaneous measurements of the flow field and wall heat transfer maps. The heat-transfer model and measurement uncertainties are also discussed briefly. Figure \ref{fig:setup} shows a schematic of the experimental setup.

The experiments were conducted in the water tunnel facility of the Department of Aerospace Engineering at the Universidad Carlos III de Madrid, with a rectangular test section of $0.5\times0.55$ m$^2$, a length of $2.5$ m, a speed ranged from $0.1$ m/s to $2$ m/s, and the free-stream turbulence intensity below $1\%$. For the experiments, the tunnel was operated in an open channel configuration, with a free-stream velocity, $U_\infty$, set at $0.24$ m/s. The turbulent boundary layer developes on a vertically-mounted flat plate spanning the full length of the test section. Turbulent-flow transition is induced by a zigzag-trip turbulator of 10 mm width and 2 mm thickness, which is mounted 120 mm downstream of the leading edge, thus 1080 mm upstream of the heat transfer sensor. The tripping is followed by a V-shape embossed tape. A full description of the characteristics of the flat plate is reported in the work by~\cite{foroozan}.

Convective heat transfer measurements were carried out using a flush-mounted heated-thin-foil sensor embedded in the wall and an IR camera used as a temperature transducer. The IR images are recorded at $59$ Hz with a FLIR SC4000 camera. The noise equivalent temperature difference (NETD) of the sensor is $18$ mK. The spatial resolution is approximately $1.1$ pixels per mm. The heat transfer sensor was installed on the flat plate at approximately $1.2$ m from the leading edge. The TBL parameters, estimated with Ensemble Particle Tracking Velocimetry as in \cite{sanmiguel2017adverse}, are reported in Table \ref{Table_TBL_char}.  

\begin{table}
\caption{Boundary layer parameters at the measurement location.}\label{Table_TBL_char}%
\begin{tabular}{@{}ll@{}}
    \toprule
    Parameter & value \\
    \midrule
    $\delta_{99}$ &   34 mm \\
     $Re_\theta$ &   1104  \\
     $Re_\tau$ &   390  \\
     $H$ &   1.5  \\
     $u_\tau$ &   0.0104 m/s  \\
    \botrule
\end{tabular}
\end{table}

The sensor, made of a thin constantan foil of $28\mu$m thickness, was heated by Joule effect, providing a constant heat flux. For a detailed description of this sensor the reader is referred to the work by ~\cite{foroozan}.  Measuring the input heat flux along with the foil temperature allows us to estimate the convective heat transfer coefficient ($h$) between the foil and the flow from the unsteady energy balance on the foil as in the work by \cite{nakamura2009frequency} and \cite{raiola2017towards}. Taking into account tangential conduction and the foil thermal inertia, the instantaneous convective heat transfer coefficient $h$ was recovered. This can be reported later in terms of Nusselt number ($Nu = h\delta_{99}/k$), where $\delta_{99}$ is the local boundary layer thickness, and $k$ is the thermal conductivity of water at wall temperature.

\begin{figure}
	\centering
	\includegraphics[width=0.7\textwidth]{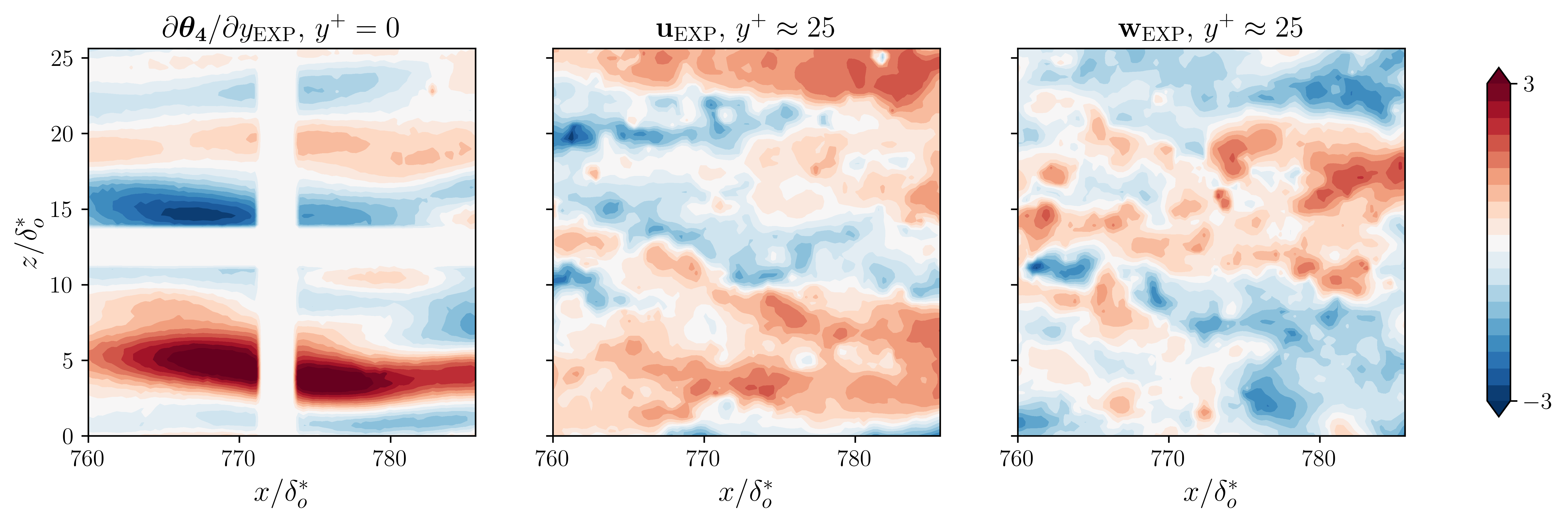}
	\caption{Heat-flux and flow-field visualizations, as sampled from the water tunnel. Note the foil support, which limits the possibility of measuring the heat-flux field.}
	\label{fig:exp_fields}
\end{figure}

Figure~\ref{fig:exp_fields} reports an example of an instantaneous wall $Nu$ field in the streamwise/spanwise plane obtained using IR thermography. The white cross corresponds to a masked region due to the presence of the foil-support structure to minimize its bending and deformation on the foil due to water pressure. Using this sensor, the Nusselt-number uncertainty was estimated to be lower than $6\%$, accounting for measurement uncertainties, following the same uncertainty-characterization process described by \cite{foroozan}.

Velocity fields were measured with wall-parallel planar PIV, in the logarithmic layer of the TBL profile centred at $y^+\approx 25$. The laser sheet thickness is approximately $1$ mm, i.e. $11$ wall units. The PIV images have been captured at a frequency equal to 1/4 of the IR acquisition one.  The PIV images have been captured with a resolution of $26.4$ pixels per mm employing an Andor Zyla sCMOS camera (with a sensor
of $2160 \times 2560$ pixels). The raw images were pre-processed to remove background reflections \citep{mendez2017pod}, and the velocity fields were evaluated using custom-made software developed at the University of Naples Federico II \citep{Astarita2004,Astarita2007}. The PIV code applies digital correlation \citep{Willert1991} with an iterative multi-grid/multi-pass algorithm \citep{SORIA1996} and image deformation \citep{scarano2001} as the interrogation strategy, with final interrogation windows of $40 \times 40$ pixels and $75\%$ overlap.

Overall, we have about 5,600 samples for training and 1,300 for validation. The test dataset includes approximately 1,100 samples.

\section{Methodology}
\label{sec:meth}
\subsection{Neural-network model}
\label{ssec:NN_model}
In this work, we consider several network architectures for different types of predictions. Based on the quantities sampled at the wall that are provided as input to the neural network, three types of predictions are investigated, as summarized in table~\ref{tab1}. 
\begin{table}
        \caption{Summary of inputs and outputs for different prediction types. The inputs are measured at the wall, the outputs are sampled at a given wall-normal location. The different types of predictions are performed for the four available Prandtl numbers.}
        \label{tab1}
        \begin{tabular}{c c c}
        \toprule
        Type & Inputs & Outputs \\
        \midrule
        I	& $\partial u/\partial y,\ \partial w/\partial y,\ p$ & $u,\ v,\ w$ \\
        II	& $\partial\theta_i/\partial y,\ \partial w/\partial y,\ p$ & $u,\ v,\ w$ \\
        III	& $\partial\theta_i/\partial y$ & $u,\ v,\ w$ \\
        \botrule
        \end{tabular}
\end{table}

In the first problem, a neural-network model is trained to predict the velocity-fluctuation fields farther from the wall using the streamwise and spanwise wall-shear-stress components, as well as the wall-pressure fields. The predictions of the first problem are denoted as \textit{type I}. These predictions use the same inputs and outputs as those of our previous work~\citep{guastoni_2021}. In the second problem, the streamwise wall-shear stress is substituted with the heat flux field corresponding to a passive scalar. We refer to these predictions as \textit{type-II} predictions. Finally, a third problem is considered, using only the heat-flux field as input (\textit{type III}). The latter two types of predictions are performed using all four Prandtl numbers sampled from the DNSs. Type III predictions aim to reproduce our experimental setting, in which we will be able to measure only the wall heat-flux field.

All the trained models are fully-convolutional neural networks (FCNs), meaning that the input information is processed by a sequence of convolutional layers, but there are no fully-connected layers at the end, as it is typical of convolutional neural networks that are employed for classification tasks on the entire input.
The inputs of the FCN model are normalized with the mean and standard deviation computed on the training samples. The velocity-fluctuation fields predicted by the FCN are scaled with the ratio of the corresponding root-mean-squared (RMS) values and the streamwise RMS value, following~\cite{guastoni_2021}. The scaled output quantities are indicated with $\widehat{\bullet}$.

FCNs allow an accurate reconstruction of the flow fields thanks to their capability to identify simple features and to combine them into progressively more complex ones. The FCN used as reference~\citep{guastoni_2021} is relatively shallow (\textit{i.e.} few convolutional layers), with a high number of kernels per layer. On the other hand, the network architectures tested in this work have a higher number of layers with fewer kernels per layer. These modifications are designed to enhance the compositional capabilities of the model without increasing its GPU-memory footprint and computational-training cost. 
Note that deeper networks can be harder to train, since they are more evidently subjected to vanishing-gradient problems. Several solutions have been proposed in the literature, including but not limited to batch normalization~\citep{batchnorm} and dropout~\citep{dropout}. We include the former in our architecture but not the latter, since dropout has been mostly used after fully-connected layers in the literature, which are not present in our models. Additionally, it should be noted that the output of each convolutional layer is slightly smaller than the input, depending on the size of the convolutional kernel~\citep{dumoulin2016guide}. When a very high number of layers is used, the output can become significantly smaller than the input. In our work, the size of the output is kept constant by modifying the size of the input field according to the architecture. This is realized by sampling a larger area in the streamwise direction and by padding periodically the field in the spanwise direction. 
Different models with a varying number of layers and parameters are trained in order to identify the best combination of these network architecture parameters. Such comparison is performed using HD DNS data, on {\it type-III} predictions at $y^{+}=30$ with $\theta_4$ heat flux as input. The objective is to optimize the network performance for its experimental use, as further detailed in section~\ref{ssec:netarch}.
Unless stated otherwise, all the considered neural-network models are trained using the hyperparameters described in table~\ref{tab:hypr}. Note that the number of samples per batch is limited only by the memory of the GPU used for training. On the same GPU, we are able to use a larger batch-size when HD samples are used.
\begin{table}
    \caption{Hyperparameters used for the training of the neural-network models.} 
    \label{tab:hypr}
    \begin{tabular}{c c}
    \toprule
    Parameter & Value \\[3pt]
    \midrule
    Learning rate & $0.001 \cdot 0.5^{\mathrm{epoch}/20}$ \\[3pt]
    Epochs & 50 \\[3pt]
    Batch size (FD) & 16 \\[3pt]
    Batch size (HD) & 32 \\[3pt]
    Batch size (experimental) & 32 \\
    \botrule
    \end{tabular}
\end{table}
The FCN is trained using the Adam~\citep{kingmaba} optimization algorithm to minimize the mean-squared error (MSE) of the predictions with respect to the turbulent fields sampled from the DNS:
\begin{equation} \label{eq:loss}
\mathcal{L}(\mathbf{\widehat{u}}_\mathrm{FCN};\mathbf{\widehat{u}}_\mathrm{DNS})=\frac{\sum_{i=1}^{N_{x,s}} \sum_{j=1}^{N_{z,s}} \left | \mathbf{\widehat{u}}_\mathrm{FCN}(i,j) - \mathbf{\widehat{u}}_\mathrm{DNS}(i,j)\right |^{2}}{N_{x,s} N_{z,s}},
\end{equation}
where boldface indicates the vectors containing the three velocity components and $|\bullet|$ represents the $L_2$ norm. We refer to the error in the individual components using $\mathcal{L}(\bullet)$ for brevity.

For {\it type-III} predictions, an additional auxiliary loss function is considered: streamwise and spanwise wall-shear stress as well as wall-pressure field are predicted by the network as an intermediate output, in an effort to drive the internal flow representation of the FCN towards physically-meaningful and interpretable quantities. Also in this case, the network parameters are updated based on the gradient with respect to the MSE between the reference DNS quantities and the FCN predictions.

\subsection{Transfer learning from synthetic experimental data}
\label{ssec:sed}
Our objective is to perform the predictions on the experimental data with the highest possible accuracy. Our dataset only includes few thousands of samples, as reported in section~\ref{ssec:exp_data} and the limited availability of data may hinder the prediction performance of the FCNs. To address this issue, we modify the DNS data in order to obtain a dataset of synthetic experimental samples. Using this data, we can optimized a neural network model and then use transfer learning~\citep{pan2009survey}. With this approach, we fine-tune the trained model on the experimental data.

In order to improve the transfer effectiveness, the synthetic data need to match the size and resolution of the experimental samples as much as possible. The experimental data have a lower resolution than their DNS counterpart, they are obtained in a relatively small region and they encompass a very limited range of Reynolds numbers. The sensor field of view is approximately equal to $0.04\,{\rm m}\cdot0.04\,{\rm m}$, which approximately translates to $24\delta^*_0 \cdot 24\delta^*_0$. We take a subset of the DNS fields of size $30\delta^*_0 \cdot 30\delta^*_0$, with $x/\delta_0^* \in [760,790]$. The Reynolds number at the center of the samples is $Re_{\tau}\approx390$, similar to the value measured in the experiments. The resulting input and output fields have a resolution of $96\times96$. Taking DNS fields of this size would result in a limited number of points per sample for the neural network training. In order to address this issue, we increased the number of samples by considering a wider streamwise range ($x/\delta_0^* \in [745,805]$), from which we take three samples of size $30\delta^*_0 \cdot 30\delta^*_0$, with $15\delta^*_0$ overlap in the streamwise direction. Since our simulation domain has a spanwise size $L_z = 50\delta^*_0$, it is possible to take two rows of samples with overlap $10\delta^*_0$. Overall, we obtain six samples from each DNS field.

While in the DNS we have heat-flux information in every point at the wall, in the experiment we need to take into account the limitations of the measurement system. The support used to limit the heat-transfer sensor deformation prevents the heat flux measurements at that location. In our synthetic data, we remove information from the input in the same way, setting to zero the heat-flux value at the points corresponding to the support.

Since the wall-normal component of the velocity fluctuations is not measured in the experiments, we train the FCN to predict only the streamwise and spanwise components. 

\section{Results and discussion}
\label{sec:results}
\subsection{Network-architecture choice and predictions on DNS data}
\label{ssec:netarch}
\begin{figure}
	\centering
    \begin{overpic}[width=\textwidth]{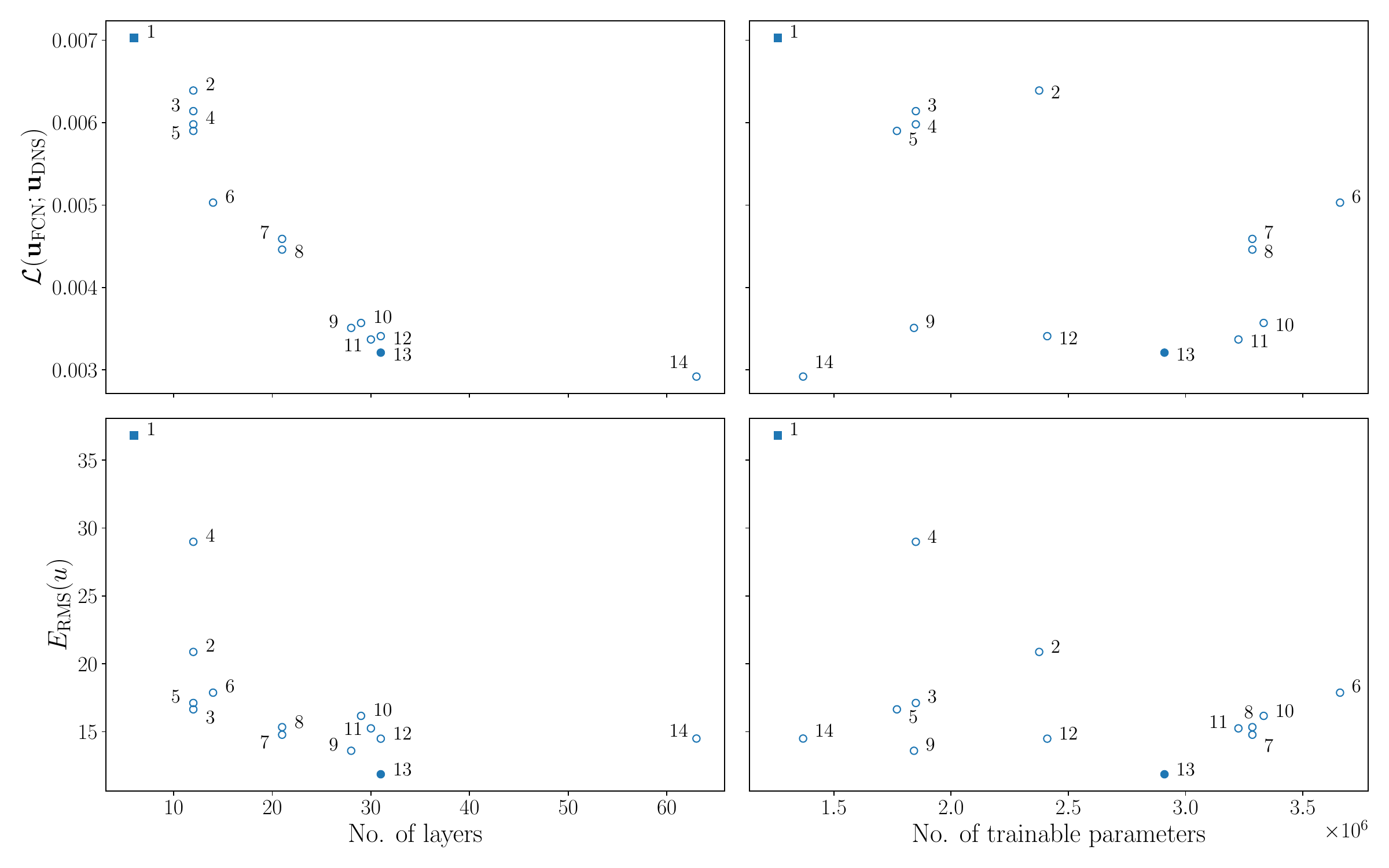}
        \put (490,570) {a)}
        \put (950,570) {b)}
        \put (490,285) {c)}
        \put (950,285) {d)}
    \end{overpic}
    \caption{MSE and RMS error in the predictions as a function of the number of layers of the network or the number of trainable parameters, for {\it type-III} predictions. $\partial\theta_4/\partial y |_{\mathrm{wall}}$ is used as input. The numbers are indexes to identify the different architectures, sorted by the number of layers. The filled square marker represents the network model used in~\cite{guastoni_2021}, while the filled circle is the model proposed in this work. These two latter models are compared in all the subsequent analysis.}
	\label{figure0}
\end{figure}

The quality of the network predictions is assessed on the test dataset, which consists of samples that are uncorrelated to the data used for training. The comparison is performed using the MSE with respect to the corresponding DNS fields and the turbulence statistics accuracy. The pre-multiplied two-dimensional power-spectral densities are also computed, to assess the amount of energy reconstructed for the different scales. 
We compare different FCN architectures with varying number of layers and trainable parameters by analyzing the performance in terms of MSE and root-mean-squared (RMS) error in the streamwise velocity fluctuations. All the models are trained to perform {\it type-III} predictions. The number of layers in the network appears to be the most impacting factor on the accuracy of the predictions, as highlighted in figure~\ref{figure0}a,c. The MSE decreases as the number of layers is increased. The error in the statistics also follows a similar trend, however, it should be noted that the deepest network considered shows a slightly higher error than the one immediately shallower. Increasing the number of trainable parameters in the network does not have a clear effect on the MSE (figure~\ref{figure0}b) and on the RMS error (figure~\ref{figure0}d). 

The network architecture trained in~\cite{guastoni_2021} and the deepest network in this work have roughly the same number of trainable parameters, however, the latter network shows a prediction error that is about 50\% lower than the former. This result suggests higher importance of the compositional capabilities of the network over its capacity. Despite achieving the best performance in the comparison, our deepest network was not selected for the subsequent analysis primarily because of the low ratio between the output and input field size. In particular, in an experimental setting, we would not be able to increase the size of the input fields as done in this numerical investigation, hence a deeper network would inevitably result in a smaller output field in which fewer turbulent features are represented. 
In the remainder of this work, we present the results obtained with the second best architecture, which has about half of the layers as our deepest one, allowing us to maintain a more acceptable output/input size ratio, while providing a comparable performance in terms of MSE. This architecture also provides a slightly smaller error in the RMS error for the streamwise fluctuations, as shown in figure~\ref{figure0}c. Note that this network has a higher number of trainable parameters, having a higher number of kernels per layer than the deepest network trained. 

\begin{figure}
	\centering
	\includegraphics[width=0.63\textwidth]{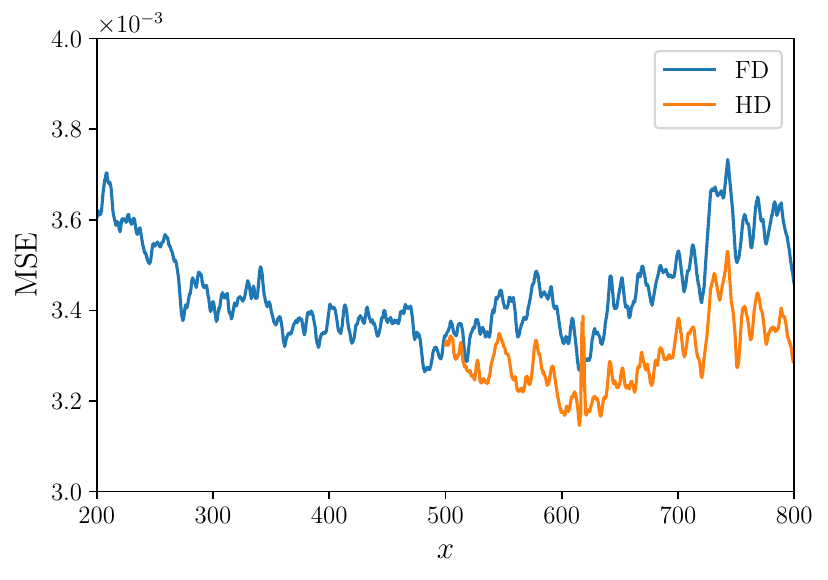}
	\caption{MSE in the predictions as a function of the streamwise location for {\it type-III} predictions at $y^+=30$. The curve represents the average over the spanwise direction and the samples in the test dataset.}
	\label{figure01}
\end{figure}

Given the network architecture, the predictions on the full-domain and on the half-domain datasets are compared. One of the advantages of the FCN is that the architecture does not depend on the size of the input. Either datasets can then be used to train the neural-network model. Despite providing more information per sample during training, the predictions of the model optimized on the full-domain dataset are less accurate than the ones on the half-domain dataset. This can be explained by considering that the boundary layer is a spatially-developing flow. This means that each sample contains a range of Reynolds numbers that need to be predicted. If we consider the error in the predictions along the streamwise direction shown in figure~\ref{figure01}, the trend exhibits a minimum at the center of the sample. This is a result of the use of MSE as training objective function. Furthermore, the larger the sample, the lower the Reynolds number at which we can achieve the highest accuracy. If we consider a smaller range encompassing only the higher Reynolds numbers (\textit{i.e.} we use HD samples), more accurate predictions can be obtained at high $Re$. 
This result is encouraging for experimental applications, since the input data that can be obtained are limited to a small interrogation window (\textit{e.g.} obtained from particle-image velocimetry), with a small Reynolds-number range. 
It should be noted, however, that training the models on a smaller portion of the domain reduces the overall amount of data available for optimization for a given size of the training dataset. 
Further shrinking of the domain size will eventually yield a performance reduction because of this trade-off
Because of these observations, the prediction results in the subsequent part of the paper will be only related to the half-domain dataset.

\begin{figure}
	\centering
	\includegraphics[width=\textwidth]{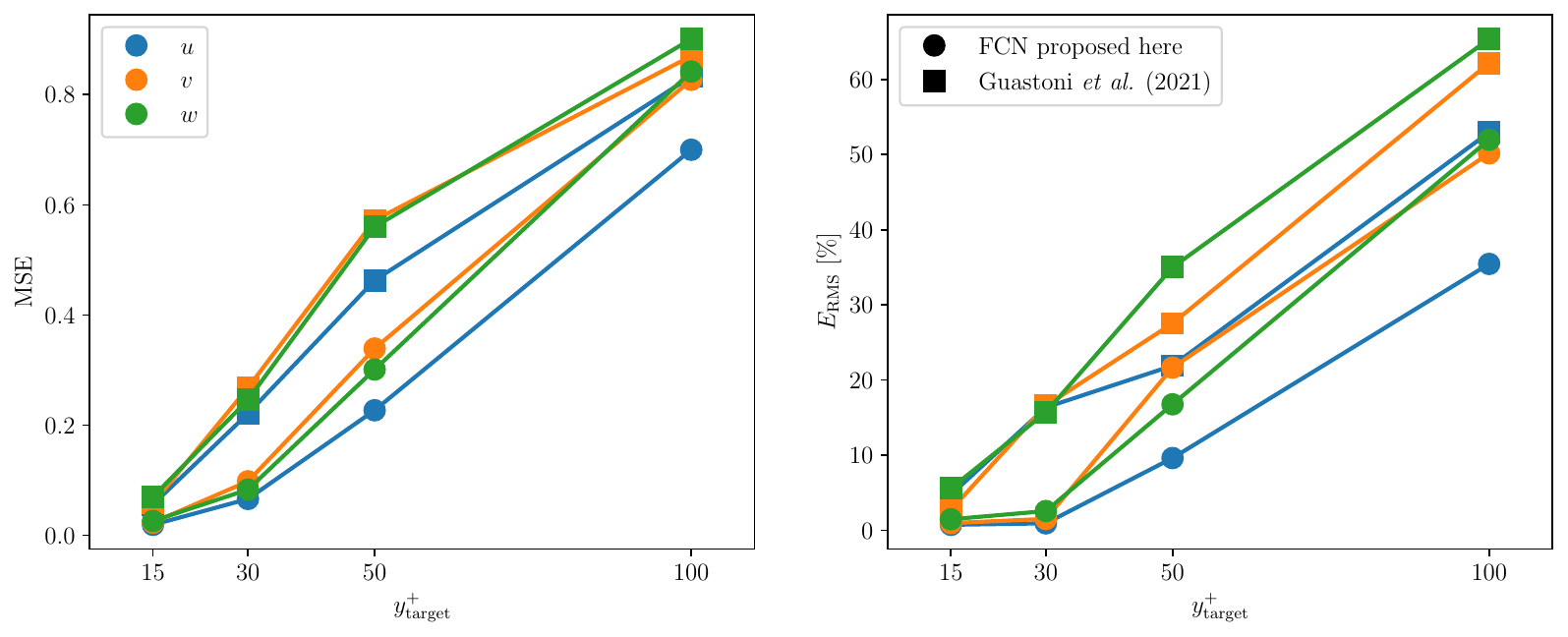}
	\caption{MSE (left) and turbulence-statistics error (right) obtained in {\it type-I} predictions with respect to target fields at different wall-normal locations. The error for each velocity component is normalized with the square of the corresponding fluctuation intensity.}
	\label{figure1}
\end{figure}
\begin{figure}
	\centering
	\includegraphics[width=0.95\textwidth]{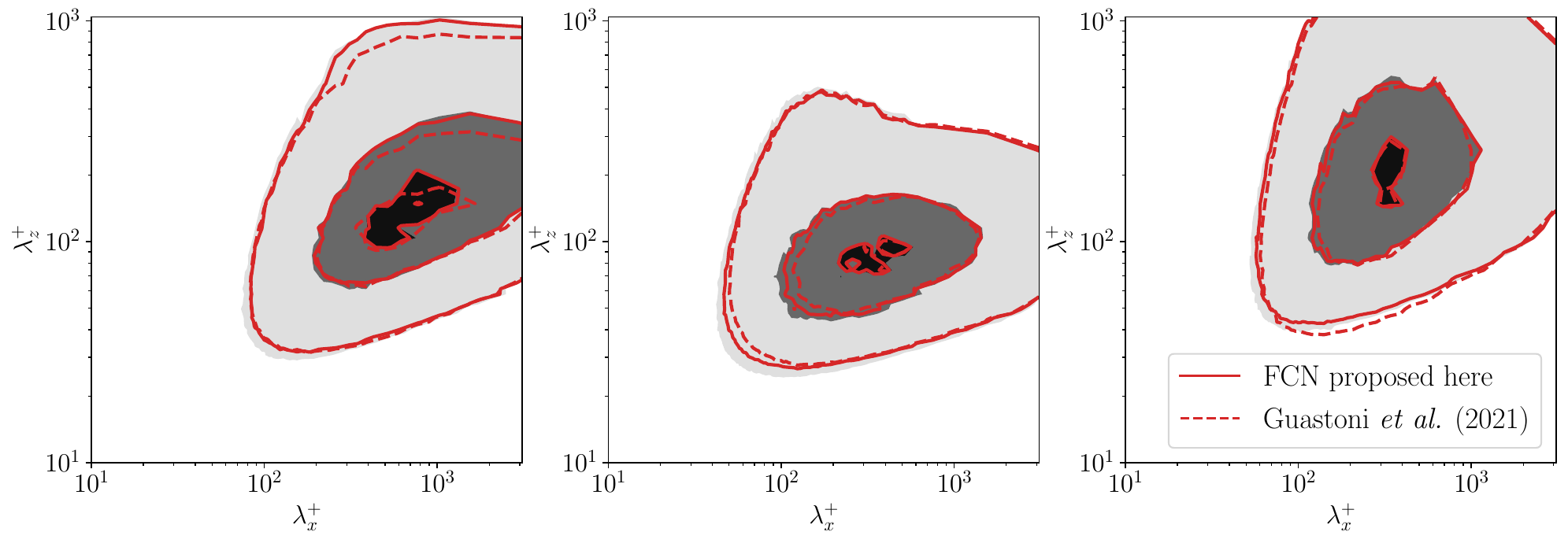}
	\caption{Pre-multiplied two-dimensional power-spectral densities for type I predictions at $y^+=30$. The three columns represent $k_{z}k_{x}\phi_{uu}$ (left), $k_{z}k_{x}\phi_{vv}$ (center), $k_{z}k_{x}\phi_{ww}$ (right). The contour levels contain 10\%, 50\% and 90\% of the maximum DNS power-spectral density. Shaded contours refer to the reference data, while contour lines refer to the FCN proposed here (solid) and the network by~\cite{guastoni_2021} (dashed), respectively.}
	\label{figure_spect}
\end{figure}
Once the network architecture and the dataset are chosen, the model is trained three times, unless noted otherwise. Each training run is performed with different random initialization in order to verify the consistency of the stochastic optimizations. The reported results show the average performance of the three models.
When three inputs are considered ({\it type-I} and {\it type-II} predictions), the present network is able to reconstruct the non-linear relation between input and output fields with higher accuracy than the FCN proposed in~\cite{guastoni_2021}, as shown in figure~\ref{figure1} ({\it type-I} predictions). The improvement is consistent across the entire range of investigated wall-normal locations. On the other hand, the performance degrades in a similar way as we move farther away from the wall. The accuracy of the present network at $y^{+}=50$ is comparable to that of the FCN in~\cite{guastoni_2021} at $y^{+}=30$. At $y^{+}=100$, the MSE in the wall-normal and spanwise directions is similar for both architectures. The accuracy improvement is even more pronounced when considering the predicted turbulent statistics. In short, the error is lower at all $y^{+}$ locations and even at $y^{+}=100$, the current FCN performs substantially better than its predecessor.
The difference in the compositional capability is evident when comparing the pre-multiplied power-spectral densities of the predictions of the two networks in figure~\ref{figure_spect}. The small-scale features at $y^+=30$ are well predicted by both networks, however, the deeper FCN can better reproduce the larger-scale features, especially in the spanwise direction.

\begin{figure}
	\centering
	\includegraphics[width=\textwidth]{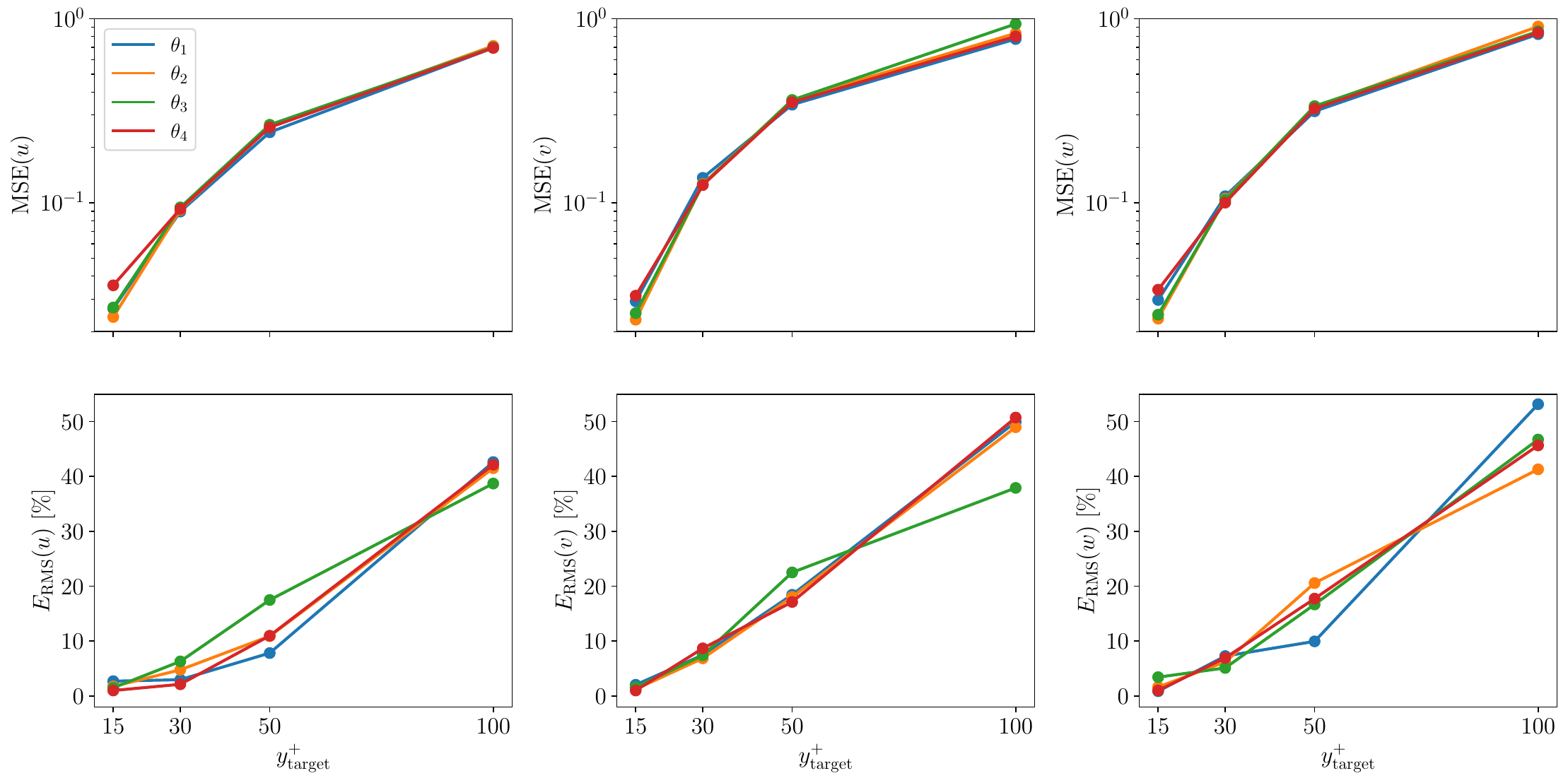}
	\caption{MSE (top) and turbulence-statistics error (bottom) obtained in {\it type-II} predictions with respect to target fields at different wall-normal locations, using different scalar fields as inputs. The error for each velocity component is normalized with the square of the corresponding fluctuation intensity.}
	\label{figure_typeII}
\end{figure}

\textit{Type-II} predictions represent an intermediate step toward the flow estimation using only the heat flux. When passing from \textit{type I} to \textit{type II}, the error is higher for all the velocity components, however, the streamwise component is the most affected, as shown in table~\ref{tab_I_vs_II}. The heat flux at the wall is less correlated to the velocity-fluctuations away from the wall than the streamwise wall-shear stress. Despite neural networks provide a non-linear mapping between input and output, a performance reduction can be anticipated by observing the linear-correlation measures in appendix~\ref{secA1} A higher Prandtl number for the scalar determines a higher error in the predictions, especially close to the wall. As shown in figure~\ref{figure_typeII}, the difference between the different scalars is less pronounced farther away from the wall. Including the spanwise wall-shear stress and the pressure as inputs makes the Prandtl dependence less evident in this type of predictions. The bottom row of figure~\ref{figure_typeII} highlights an apparent inconsistency in the relation between the statistical error and the Prandtl number. Since we do not train the model for statistical accuracy directly, the error has a higher variance between models than the MSE on the individual predictions. Furthermore, only one model per Prandtl number and wall-normal location has been trained in this case. Averaging the results from different models helps retrieving the relation between the error and the Prandtl number that can be observed in the MSE. The percentual increment of the MSE due to the added difficulty of the predictions is similar for both the network proposed in this work and the one  in~\cite{guastoni_2021}; note however that the predictions from the former are significantly more accurate than those of the latter. For this reason, the current FCN is found to perform better across the entire range of wall-normal locations, even for \textit{type-II} predictions.
\begin{table}
    \caption{Normalized MSE comparison for the different velocity components in \textit{Type-I} and \textit{Type-II} predictions at $y^{+} = 15$. \textit{Type-II} predictions results are shown using different scalar fields as inputs.} 
    \label{tab_I_vs_II}
    \begin{tabular}{c c c c c c}
    \toprule
    & Type I & Type II ($\theta_1$) & Type II ($\theta_2$) & Type II ($\theta_3$) & Type II ($\theta_4$)\\[3pt]
    \midrule
    $\mathcal{L}(u)/u^2_\mathrm{RMS}$ & 0.019 & 0.027 & 0.024 & 0.027 & 0.036 \\[3pt]
    $\mathcal{L}(v)/v^2_\mathrm{RMS}$ & 0.023 & 0.029 & 0.023 & 0.025 & 0.031 \\[3pt]
    $\mathcal{L}(w)/w^2_\mathrm{RMS}$ & 0.026 & 0.030 & 0.024 & 0.025 & 0.034 \\
    \botrule
    \end{tabular}
    
\end{table}

\begin{figure}
	\centering
	\includegraphics[width=\textwidth]{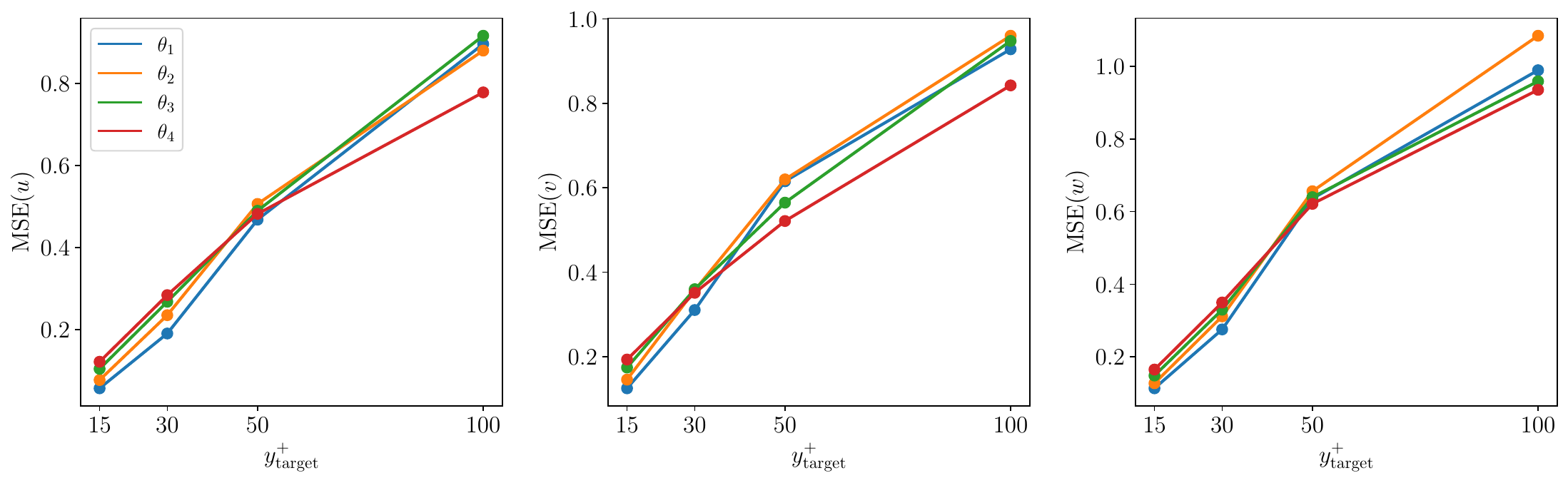}
	\caption{MSE obtained in {\it type-III} predictions with respect to target fields at different wall-normal locations, using different scalar fields as inputs. The error for each velocity component is normalized with the square of the corresponding fluctuation intensity.}
	\label{figure_typeIII}
\end{figure}

When {\it Type-III} predictions are performed, the resulting MSE is about three times higher when compared to {\it type-II} predictions, confirming that information in the spanwise wall-shear stress and wall pressure has an important role in the reconstruction of the fields away from the wall. Figure~\ref{figure_typeIII} shows how the use of a different scalar does affect the prediction quality. Close to the wall, the predictions become progressively more challenging as the Prandtl number increases. Interestingly, the difference is negligible at $y^+=50$ and the models trained at $y^+=100$ with $Pr=4$ and $Pr=6$ perform better than the models trained with lower-Prandtl-number inputs. Note, however, that the error in all these reconstructions is quite significant and only the largest structures are predicted. The degradation in the prediction quality with respect to the wall-normal distance was already reported in~\cite{guastoni_2021}.

\begin{figure}
	\centering
	\includegraphics[width=0.47\textwidth]{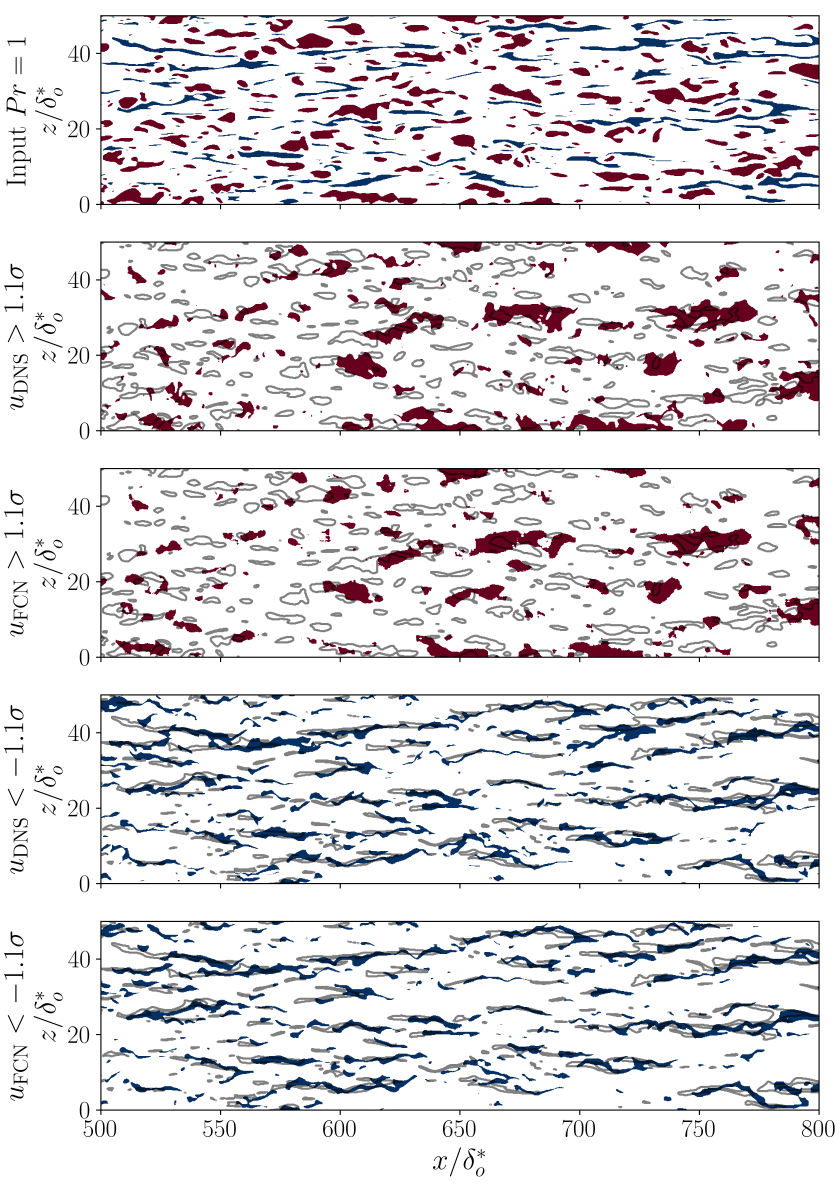}
    \includegraphics[width=0.47\textwidth]{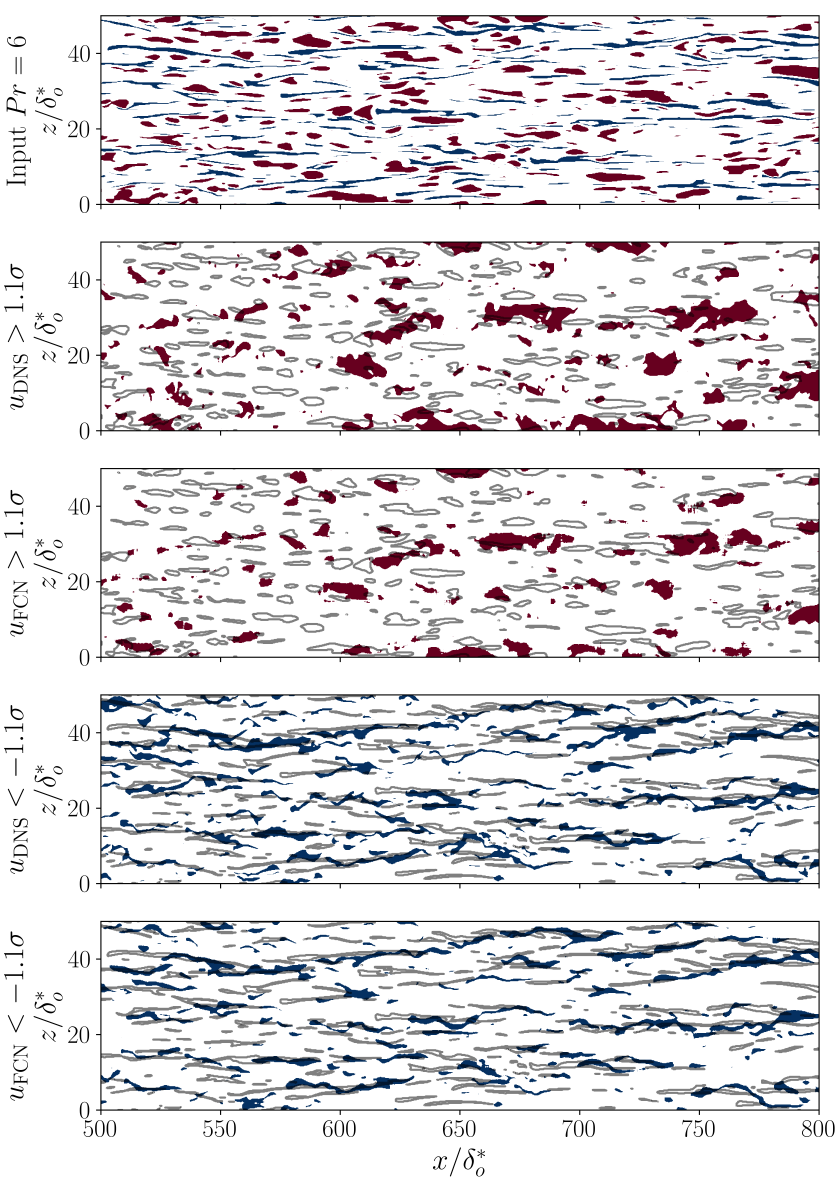}
	\caption{Comparison of the regions of highest and lowest fluctuations in the input fields (top row), and the output fields at $y^+=30$ from DNS and FCN prediction. The second and third rows show the highest fluctuations in the DNS and FCN prediction, respectively. The fourth and fifth rows show the lowest fluctuations. The grey contours represent the corresponding positive or negative fluctuations in the input field.
    The left column shows a sample {\it type-III} prediction using the heat flux at $Pr=1$. The right column shows the predictions with input at $Pr=6$. Blue and red indicate the regions where the fluctuations are smaller than $-1.1\sigma$ and larger than $1.1\sigma$, respectively.}
	\label{figure_pr_vs_pr}
\end{figure}

Note that, closer to the wall, the predictions become more and more difficult as the Prandtl number increases. This can be linked with the change of the heat-flux features that can be observed by the neural network as the Prandtl number changes. When the mean-squared error is used to optimize the neural network, the predicted fields tend to span a smaller range of values than the corresponding DNS fields. The correct localization and estimation of the strongest fluctuations becomes very relevant to obtain accurate predictions. Figure~\ref{figure_pr_vs_pr} compares the highest and lowest fluctuations of the input and output fields at $y^+=30$ for two different Prandtl numbers. At $Pr=1$, the features in the output have a spatial correspondence with the features in the input, and this is particularly evident for the lowest fluctuations. Since the neural network provides a localized relation between input and output, an accurate reconstruction is possible. 
The spacing between the higher and lower values in the input becomes smaller in the spanwise direction as the Prandtl number increases. In this case, the prediction becomes more challenging because a larger number of high/low-value streaks are present in the receptive field of the network and the spatial location of the output features is more difficult to identify.

\begin{figure}
	\centering
        \begin{overpic}[width=0.275\textwidth]{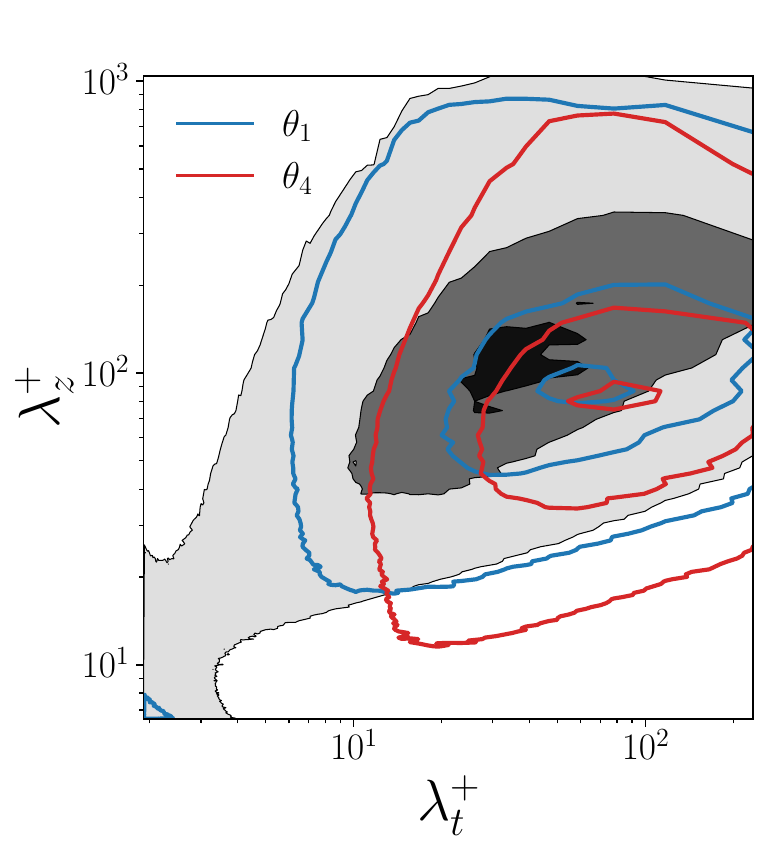}  
         \put (2,900) {a)}
        \end{overpic}
        \begin{overpic}[width=0.7\textwidth]{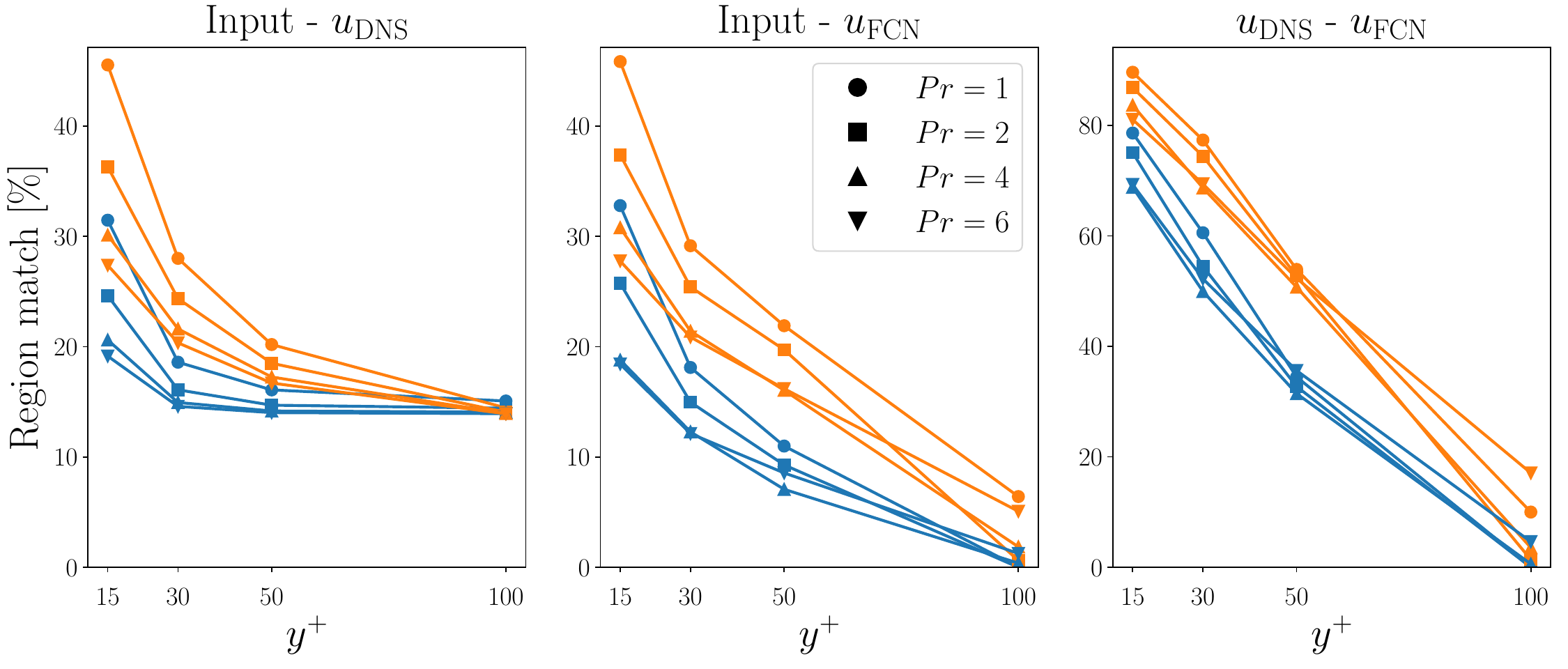}
         \put (2,390) {b)}
        \end{overpic}
	\caption{a) Pre-multiplied two-dimensional power-spectral densities of wall-heat flux and streamwise velocity fluctuation at $y^+=30$ (shaded contours), for the data-points that have $\|q\|>1.1\sigma$, where $q = \{\partial \theta_1 / \partial y,~\partial \theta_4 / \partial y,~u\}$. The contour levels contain 10\%, 50\% and 90\% of the maximum DNS power-spectral density.  b) Percentage match in the highest (blue) and lowest (orange) fluctutations between input and DNS fields (left), input and FCN predictions (center), DNS fields and FCN predictions (right). The comparison is performed for all the sampled $Pr$ numbers.}
	\label{figure_match}
\end{figure}

In order to obtain a more quantitative perspective on the previous observations, we compute the spectra of the filtered input and output fields, as shown in figure~\ref{figure_match}a. We identify the regions of the flow with high (resp. low) fluctuations by considering only the points where the value is above (resp. below) a threshold of $1.1\sigma$ (resp. $-1.1\sigma$), where $\sigma$ is the standard deviation of the considered quantity, as computed from the training dataset. 
Indeed, the strongest fluctuations at the higher Prandtl number exhibit a shift towards shorter spanwise wavelengths, with a reduced superposition to the output frequency content, when compared to lower Prandtl numbers. Note that these observations are aligned with the power-spectral-density analysis reported in~\cite{Balasubramanian_Guastoni_Schlatter_Vinuesa_2023}.
In figure~\ref{figure_match}b, we compare the region match as the percentage of points of high/low fluctuations that are present in both the fields considered. The region match between the heat-flux field and the DNS velocity field is higher for the lowest fluctuations than the highest, as observed in figure~\ref{figure_pr_vs_pr}. The percentage match reduces as we move farther away from the wall: this is expected as the flow features close to the wall are similar to their wall footprint. Note, however, that the match does not go below 15\% even at $y^+=100$. 
The effect of the Prandtl number is very evident close to the wall, the higher the Prandtl number, the smaller is the region match: for instance, we can observe a match reduction of about 20\% between $Pr=1$ and $Pr=6$ at $y^+=15$.  
The region match between the heat flux field and the FCN predictions is similar close to the wall but as we move farther away, the match is smaller. This is particularly evident in the highest-fluctuation region. Differently from the DNS fields, the match farther away from the wall gets progressively smaller, reaching near-zero value at $y^+=100$. The increase in the $Pr$ number is associated with a reduction of the region match. Finally, the match between the DNS fields and the corresponding predictions quantifies the error in the predictions at different wall-normal locations and Prandtl numbers, separating the contributions of the highest and lowest fluctuations. The regions of lowest fluctuations are better predicted at all wall-normal locations, so the high-fluctuation regions have a larger contribution to the error in the FCN predictions.

\begin{sidewaystable}
    \caption{Error comparison in \textit{Type-III} predictions at $y^{+} = 30$ using the network model from~\cite{guastoni_2021} and the one proposed here. The variance of the statistical error is computed across the different training runs.}
    \label{tab2}
    \begin{tabular}{c c c c c c c}
    \toprule
    & Guastoni \textit{et al.} & FCN & FCN & FCN & FCN & FCN \\
    &  & ($\theta_1$, no aux) & ($\theta_2$, no aux) & ($\theta_3$, no aux) & ($\theta_4$, no aux) & ($\theta_4$, aux)
    \\[3pt]
    \midrule
    $\mathcal{L}(u)/u^2_\mathrm{RMS}$ & 0.592 & 0.191 & 0.235 & 0.268 & 0.284 & 0.271 \\[3pt]
    $\mathcal{L}(v)/v^2_\mathrm{RMS}$ & 0.638 & 0.311 & 0.357 & 0.360 & 0.351 & 0.335 \\[3pt]
    $\mathcal{L}(w)/w^2_\mathrm{RMS}$ & 0.850 & 0.276 & 0.312 & 0.330 & 0.350 & 0.330 \\[3pt]
    $E_{\mathrm{RMS}} (u)~[\%] $ & 36.82 $\pm$ 0.67 & 7.43 $\pm$ 2.27 & 7.47 $\pm$ 0.90 & 13.61 $\pm$ 1.41 & 13.64 $\pm$ 2.05 & 11.87 $\pm$ 0.49 \\[3pt]
    $E_{\mathrm{RMS}} (v)~[\%] $ & 39.48 $\pm$ 0.92 & 14.56 $\pm$ 2.39 & 13.16 $\pm$ 2.98 & 20.42 $\pm$ 1.84 & 22.09 $\pm$ 2.03 & 14.14 $\pm$ 0.93 \\[3pt]
    $E_{\mathrm{RMS}} (w)~[\%] $ & 57.26 $\pm$ 1.24 & 12.31 $\pm$ 2.25 & 13.23 $\pm$ 2.18 & 18.28 $\pm$ 1.04 & 20.35 $\pm$ 3.25 & 14.04 $\pm$ 0.74 \\
    \botrule
    \end{tabular}
\end{sidewaystable}

\begin{figure}
	\centering
	\includegraphics[width=0.9\textwidth]{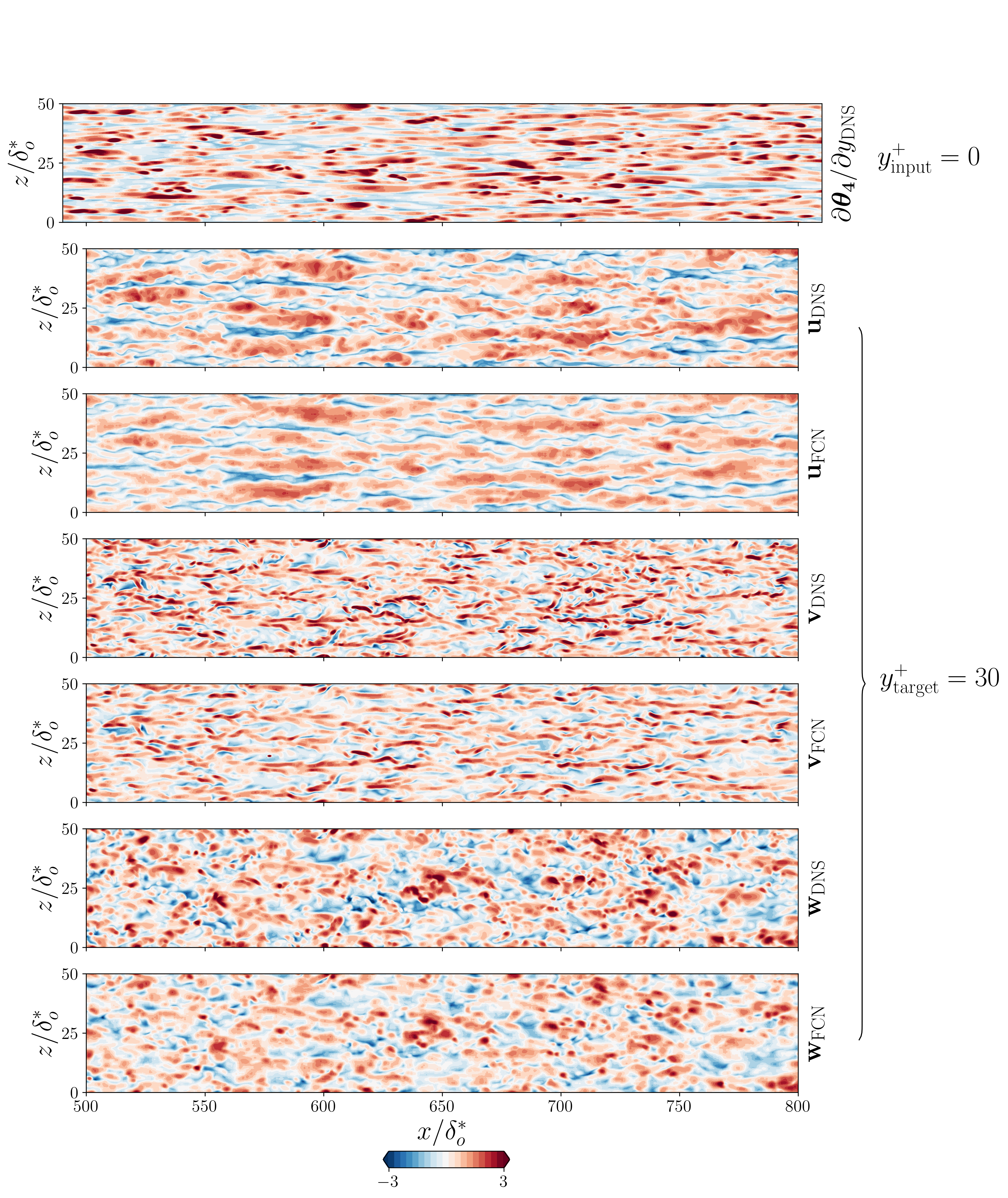}
	\caption{Sample result for {\it type-III} prediction at $y^{+} = 30$, obtained using the proposed FCN with auxiliary loss functions. The first row corresponds to the DNS input heat flux for $Pr = 6$, normalized with the mean and standard deviation computed on the training samples. The second and third rows show the streamwise DNS velocity-fluctuation field and the corresponding prediction obtained from FCN, respectively. Similarly, the fourth and fifth rows represent the wall-normal velocity fluctuation of the target and predicted fields. Finally, the sixth and seventh rows show the spanwise velocity fluctuation component of the target and predicted fields, respectively. The velocity-fluctuation fields are scaled by the respective RMS quantities.} 
	\label{figure2}
\end{figure}

The statistical error for \textit{type III} predictions can be further improved with the use of the auxiliary loss. We tested it for the predictions using $\partial\theta_4/\partial y |_{\mathrm{wall}}$ as input: while the MSE is only about $5\%$ lower, the predicted turbulence statistics are up to $20\%$ better than when the auxiliary loss function is not used, as shown in table~\ref{tab2}. The use of an auxiliary loss function also reduces the variance between models in the statistical of all the velocity components. The error comparison with~\cite{guastoni_2021} shows how a deeper FCN is necessary to achieve satisfactory predictions of this type. A sample {\it type-III} prediction at $y^{+} = 30$ using the auxiliary loss function is shown in figure~\ref{figure2}, when $\partial\theta_4/\partial y |_{\mathrm{wall}}$ is used as input. From this figure, it is possible to observe that the FCN is able to reconstruct the large-scale features of the flow in all three velocity components starting from the heat-flux field only. The smaller-features reconstruction is less accurate, in particular, the maximum positive and negative fluctuations are typically underestimated. This is related to the use of the mean-squared error as loss function for the optimization.

\begin{figure}
	\centering
	\includegraphics[width=0.95\textwidth]{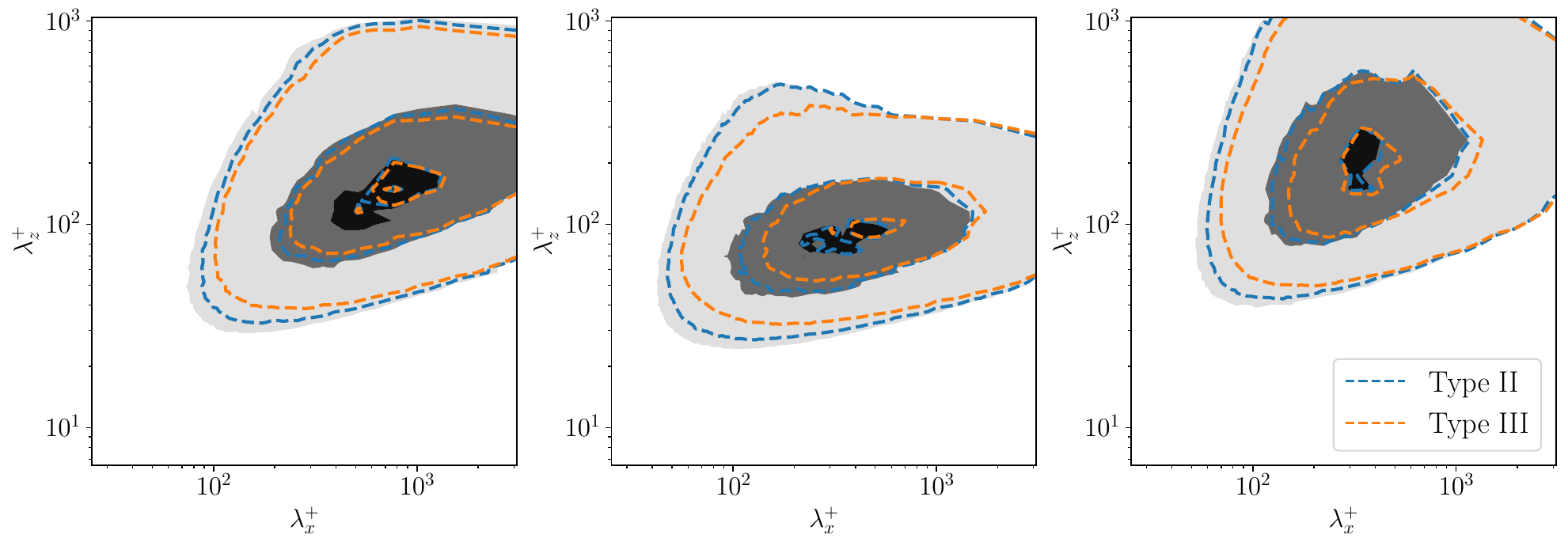}
	\caption{Pre-multiplied two-dimensional power-spectral densities of different prediction types using $\partial\theta_4/\partial y |_{\mathrm{wall}}$ as input. The three columns represent $k_{z}k_{x}\phi_{uu}$ (left), $k_{z}k_{x}\phi_{vv}$ (center), $k_{z}k_{x}\phi_{ww}$ (right). The contour levels contain 10\%, 50\% and 90\% of the maximum DNS power-spectral density. Shaded contours refer to the reference data, while contour lines refer to \textit{type-II} (blue) and \textit{type-III} (orange) predictions, respectively.}
	\label{fig:spectra}
\end{figure}

We now focus on the predictions whose setup is similar to the experimental setting. For this, we analyze more in depth \textit{type III} predictions using the heat flux at $Pr = 6$. A more comprehensive overview of the predicted energy at the different scales is provided by the spectra, shown in figure~\ref{fig:spectra}. We compare the spectra of \textit{type-II} and \textit{type-III} predictions at $y^+=30$. The amount of reconstructed energy is lower in \textit{type-III} predictions than in \textit{type-II}. Furthermore, it is possible to observe that eliminating the spanwise wall-shear stress and wall pressure has a higher impact on the prediction of the shorter wavelengths, both in the streamwise and spanwise direction. The accuracy reduction is more evident in the pre-multiplied wall-normal and spanwise spectra. This is expected, as the wall-pressure is well correlated with the wall-normal component of the velocity and the spanwise wall-shear stress helps to improve the prediction of the corresponding velocity-fluctuation component.   

\subsection{Predictions on experimental data}
\begin{figure}
	\centering
    \begin{overpic}[width=0.47\textwidth]{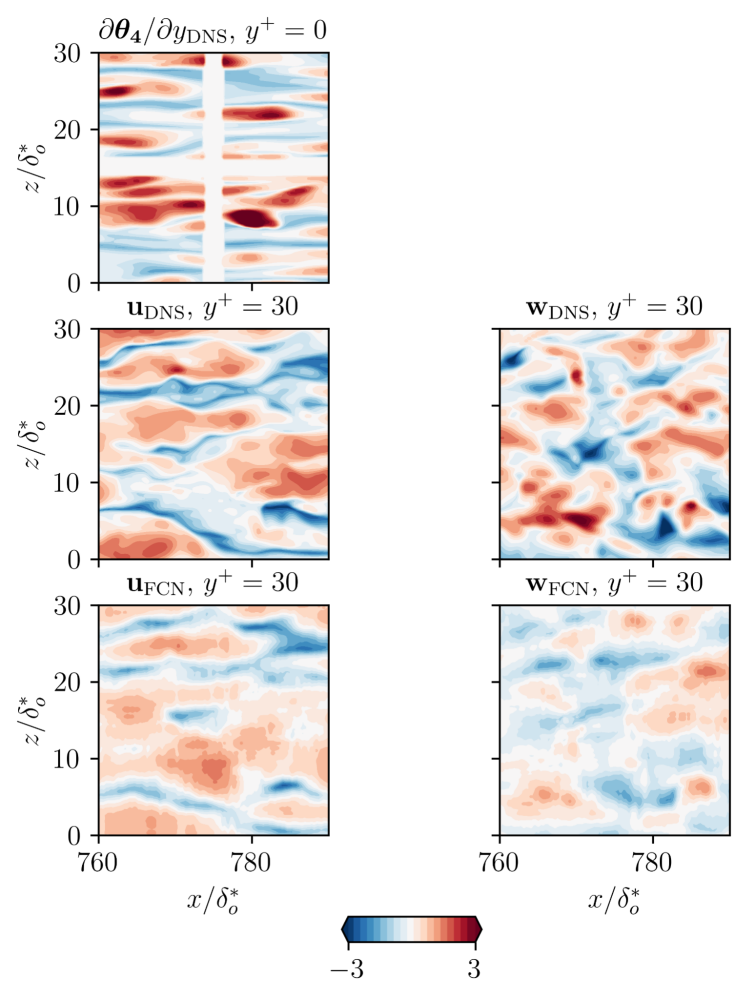}
        \put (-20,950) {\Large a)}
    \end{overpic}

    \hfill

	\begin{overpic}[width=0.47\textwidth]{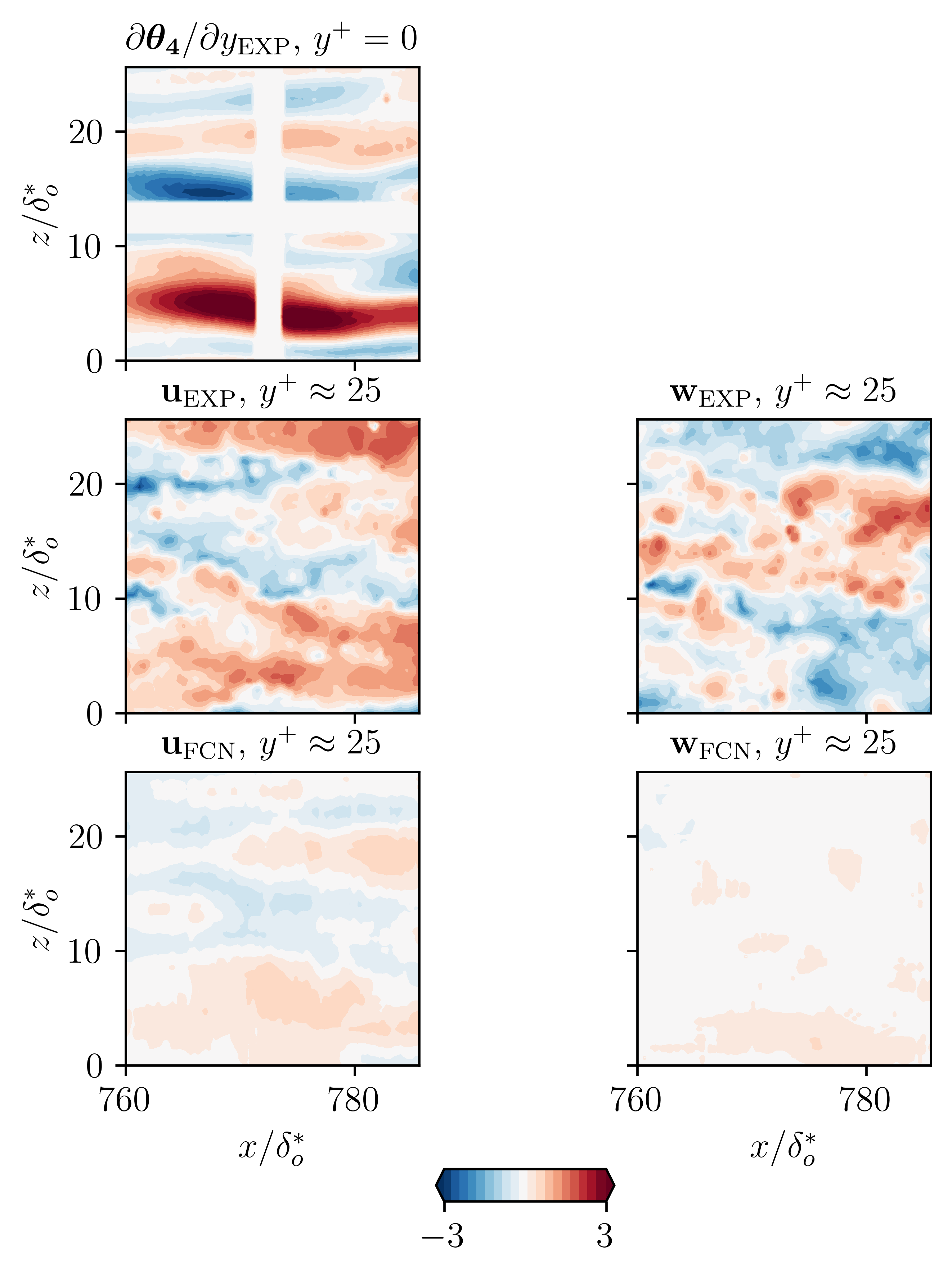}
        \put (-20,950) {\Large b)}
    \end{overpic}
	\caption{Sample result of predictions obtained using the FCN with auxiliary loss functions, trained on (a) DNS data and (b) experimental data. In both panels, the top row represents the input, \textit{i.e.} the wall heat flux field at $Pr\approx6$, the second represents the ground truth from the simulation and the experiment, respectively. Finally, the third row shows the corresponding network predictions.}
	\label{fig:exp_pred}
\end{figure}

In order to train a network on the experimental data, we first need to optimize a copy of the same neural network on the DNS data and then perform transfer learning, as described in section~\ref{ssec:sed}. 
In the previous section we selected the network architecture based on trade-off between the model performance and the input/output ratio. Note, however, that the fields sampled from the experiment are relatively small ($82 \times 82$) and they cannot be extended. For this, we have to resort to zero-padding the fields, even if this comes at the cost of a larger error close to the edges.
Also for these predictions, the performance is assessed using the mean-squared error with respect to the ground truth velocity-fluctuations fields. The use of the synthetic experimental data for training determines an increase of the MSE in the predictions with respect to the FCN models trained with the full-fields from DNS, as reported in table~\ref{tab_exp}. The error in the wall-parallel velocity-fluctuation components is more than two times larger with respect to the values reported in table~\ref{tab2}. Figure~\ref{fig:exp_pred}(a) allows a qualitative comparison of the DNS fields and the corresponding predictions. Only the larger scales of the turbulent motions are reconstructed and we can observe the prediction of regions of large positive fluctuations that are not present in the DNS field. These regions are typically located close to the foil support, where the lack of information makes the predictions harder. The streamwise component is predicted with higher accuracy than the spanwise component.

Once the FCN is fine-tuned on the experimental dataset, it is possible to assess its performance on the experimental test data. In this case, the error is about 50\% higher than the predictions on the synthetic data from DNS, as reported in table~\ref{tab_exp}. The input fields show a much smoother variation of the heat flux, as shown in figure~\ref{fig:exp_pred}(b). This is to be addressed to the spatial modulation of the heat flux sensor. The resulting predictions are less detailed than all the previous ones: in the streamwise component, a general reconstruction of the larger scales of the flow is provided. By contrast, the spanwise component reconstruction is lacking.
After gradually increasing the difficulty of the prediction, first by reducing the number of inputs and then using less and less informative ones, we are now testing the FCN network on a very challenging task. 

\begin{figure}
	\centering
    \includegraphics[width=0.53\textwidth]{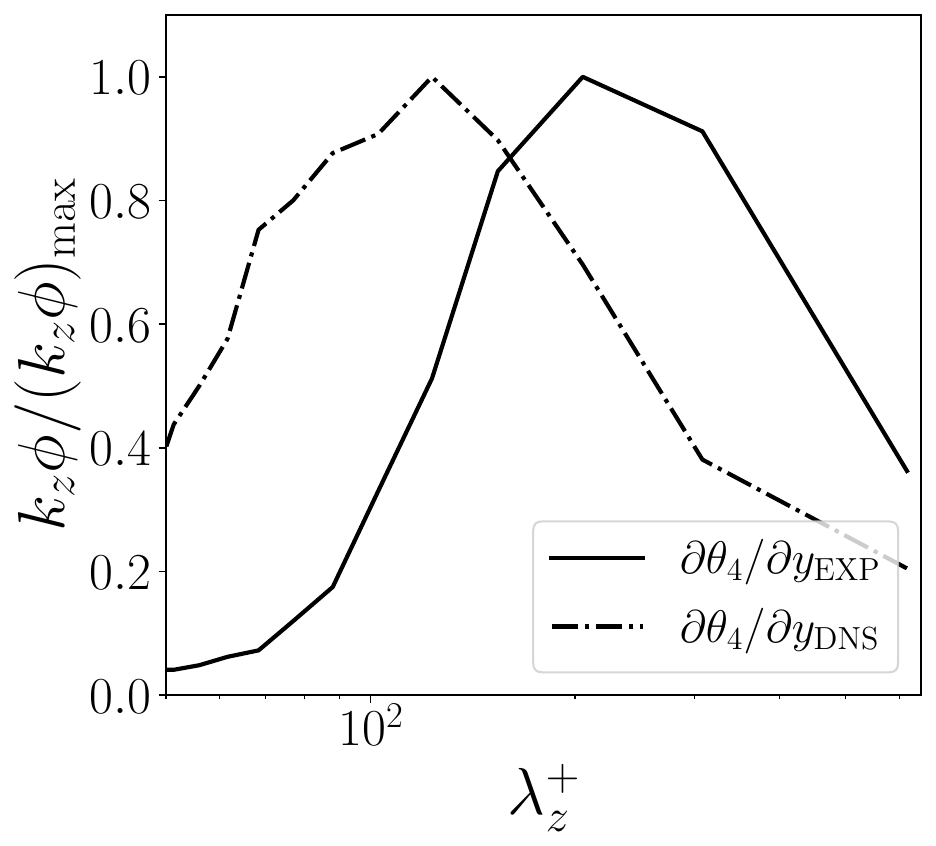}
	\caption{Comparison of the pre-multiplied one-dimensional power-spectral densities of the wall-heat flux in the synthetic DNS fields and the experimental ones.}
	\label{exp_data_comp}
\end{figure}

\begin{table}
    \caption{Normalized MSE comparison for the wall-parallel velocity components in the predictions of DNS and experimental fields.} 
    \label{tab_exp}
    \begin{tabular}{c c c}
    \toprule
    & DNS data & Experimental data \\[3pt]
    \midrule
    $\mathcal{L}(u)/u^2_\mathrm{RMS}$ & 0.587 & 0.873 \\[3pt]
    $\mathcal{L}(w)/w^2_\mathrm{RMS}$ & 0.790 & 1.000 \\
    \botrule
    \end{tabular}
    
\end{table}
The fine-tuning on the experimental training dataset can only partially compensate for the small amount of data available: the fact that the error in the prediction is higher than the DNS synthetic fields, reported in table~\ref{tab_exp}, is expected. In these conditions, the reconstruction performance of the neural network is relatively limited. These results partially depend on the reduced amount of data, but it can also be explained by the discrepancies between simulation settings and experimental conditions. In figure~\ref{exp_data_comp}, we compare the power-spectral densities of the inputs: because of the inherent limitations of the experimental procedures, the energy content at the higher frequencies is reduced. 
Furthermore, the choice of the loss function plays a role also in this context: the minimization of the MSE tends to smooth out the peaks in the predictions. For zero-mean fields, this translates to an effective reduction of the value range in the predictions.
Despite the limitations, this study represents the first attempt to perform these type of predictions using experimental data. As such, it represents an important step towards the implementation of non-intrusive sensing in experiments. Our analyses describe most of the important challenges related to the deployment of a neural network model in an engineering context, especially when the training is performed on synthetic data rather than real ones, which are often very expensive or difficult to obtain.

\section{Conclusions}
\label{sec:conc}
In this work, we assessed the prediction capabilities of a fully-convolutional network (FCN) using DNS data sampled from a turbulent boundary layer flow, with a maximum Reynolds number of $Re_{\tau}=396$. We optimize the architecture of the FCN in order to minimize the MSE in the predictions, while maintaining a satisfactory output/input ratio of the field size. For \textit{type II} and \textit{type III} predictions, we describe the effect of using as input heat flux fields at progressively higher Prandtl number. Closer to the wall, the predictions become more challenging as the Prandtl number is increased. 
With the highest Prandtl number tested ($Pr=6$), the resulting network yields {\it type-III} predictions at $y^{+} = 30$ with an error that is 50\% lower than that of the previously-studied architectures~\citep{guastoni_2021}. A higher number of layers determines a larger receptive field for the network, and it enhances its compositional capabilities. The prediction accuracy is proven to be more sensitive to this parameter than the network capacity (\textit{i.e.} the number of trainable parameters). The use of alternative, yet similar network architectures, \textit{e.g.} ResNet by~\cite{resnet} or UNet by~\cite{unet} was only partially explored and for this reason not reported, preventing a more comprehensive analysis of the available network architectures.  
The performance of completely different models, such as transformers~\citep{vaswani} and diffusion models~\citep{diffusion} could also be compared with our FCN, once they have been adapted for the described prediction task.
The architectural improvements described here are essential to achieve a satisfactory velocity field reconstruction, given the additional difficulties related to the choice of a spatially-evolving flow, and the use of input quantities that are less informative than the ones used in the previous studies. In this study we limit ourselves to the \textit{a-posteriori} analysis of the results of several trained networks, highlighting the need for deeper networks in order to obtain a satisfactory accuracy in the reconstruction. The number of layers in a network, as well as the number of trainable parameters per layer (\textit{e.g.} the number of kernels per convolutional layer) are hyperparameters that need to be tuned to obtain the best performance of a neural network model. Automated hyperparameter searches can be conducted using evolutionary algorithms or Bayesian optimization, however, several networks still need to be trained depending on the number of hyperparameters that need to be adjusted.
It should be noted that we focus our attention on improving the network performance when the heat flux is chosen as input instead of the wall-shear stress components and the wall pressure, but we have not explored possible solutions to maintain the prediction accuracy as the Prandtl number increases. In this regard, the design of a \textit{Pr-invariant} neural network model represents an appealing research direction, in order to perform accurate predictions at even higher Prandtl numbers.

After assessing the prediction capabilities of the FCN using full DNS data, we trained the selected architecture with synthetic experimental data, obtained by modifying the samples from the numerical simulation to resemble the measurements from a water tunnel. The FCN is then further optimized using samples from the experiments through transfer learning. These predictions represent the first attempt towards non-intrusive-sensing applications in experimental settings through deep learning. 
Despite the limitations of the experimental measurements, the neural-network model is able to provide acceptable predictions for the streamwise fluctuations.
It is important to highlight that the receptive field of the FCN is large compared to the size of the experimental samples. This, combined with the use of zero-padding, may have a detrimental effect on the model performance, as shown in appendix~\ref{secA2}. The network architecture can be further optimized by taking these elements into account.
The predicted fields exhibit an error with respect to the original fields that is 50\% larger than the model trained and tested only on synthetic experimental data. Future work will be devoted to improving the training procedure to reduce the performance gap between the two models.

\backmatter





\bmhead{Acknowledgements}

The authors acknowledge the Swedish National Infrastructure for Computing (SNIC) for providing the computational resources by PDC, used to carry out the numerical simulations.

\section*{Declarations}

\begin{itemize}
\item \textbf{Funding} This work is supported by the founding provided by the Swedish e-Science Research Centre (SeRC), ERC grant no.~"2021-CoG-101043998, DEEPCONTROL" and the Knut and Alice Wallenberg (KAW) Foundation.
S.D., A.I. and F.F. acknowledge funding by the project ARTURO, ref. PID2019-109717RB-I00/AEI/10.13039/501100011033, funded by the Spanish State Research Agency and the project EXCALIBUR (Grant No PID2022-138314NB-I00), funded by MCIU/AEI/10.13039/501100011033 and by ‘ERDF A way of making
Europe’.
\item \textbf{Conflict of interest} The authors report no conflict of interest.
\item \textbf{Code availability} The data that support the findings of this study are openly available in \url{https://github.com/KTH-FlowAI} upon publication
\item \textbf{Author contribution} The study was conceptualized and planned by AI, SD, and RV. The dataset was generated by AGB, with input from LG, PS and RV. The network model was implemented and trained by AGB, starting from the previous code implementation by LG.
LG, RV and HA provided feedback on the network design. The post-processing of the results was performed by AGB, with input from LG, RV and HA. The paper was written by LG and AGB. 
FF, AG, SD and AI designed the experiment. FF performed the experimental campaign with support of SD and AI. FF and AG processed the experimental data with input from SD and AI.
All authors revised the manuscript and agree with its content.
\end{itemize}







\begin{appendices}

\section{Linear correlation of inputs and outputs}\label{secA1}

Neural-network models such as FCNs are optimized to approximate the relation between input and output using a non-linear function. This capability provides a fundamental advantage with respect to techniques such as extended proper orthogonal decomposition~\citep{boree2003extended}, allowing for a better reconstruction, in particular when the output plane is far from the wall~\citep{guastoni_2021}. Nonetheless, the analysis of the linear correlation between different input-output pairs can provide a quantitative perspective on the difficulty of the individual prediction tasks. We compute the linear correlation between two variables $r$ and $s$ as:  
\begin{equation}
    R_{rs} = \langle (r-\langle r \rangle)(s-\langle s \rangle) \rangle / (r_{\rm RMS} s_{\rm RMS}),
\end{equation}
where $\langle \bullet \rangle$ indicates the average over the samples and in the homogeneous directions. Table~\ref{tab_corr} lists the correlation of the inputs--streamwise wall-shear stress and the heat fluxes at different $Pr$ numbers--and the outputs, \textit{i.e.} the velocity fluctuations in the three components. The streamwise shear stress and the heat flux at $Pr=1$ show a similar linear correlation with the velocity components. These quantities are the most correlated among all the possible inputs considered. The linear correlation decreases when the Prandtl number is higher, highlighting how predicting the velocity field based on the heat flux information at the wall represents a more and more challenging task as the $Pr$ number increases.

\begin{table}
    \caption{Linear correlation coefficient between different input-output pairs. Inputs are measured at the wall.} 
    \label{tab_corr}
    \begin{tabular}{c c c c c c}
    \toprule
    & $\partial u / \partial y$  & $\partial \theta_1 / \partial y$ & $\partial \theta_2 / \partial y$ & $\partial \theta_3 / \partial y$ & $\partial \theta_4 / \partial y$ \\[3pt]
    \midrule
    $u(y^+=15)$ & 0.516 & 0.515 & 0.386 & 0.285 & 0.240 \\[3pt]
    $v(y^+=15)$ & -0.379 & -0.413 & -0.327 & -0.245 & -0.204 \\[3pt]
    $w(y^+=15)$ & -0.0007 & -0.001 & -0.004 & -0.006 & -0.007 \\
    \botrule
    \end{tabular}
\end{table}


\section{Effect on the prediction accuracy of the padding in the convolutional layers}\label{secA2}

\begin{figure}
	\centering
	\includegraphics[width=0.67\textwidth]{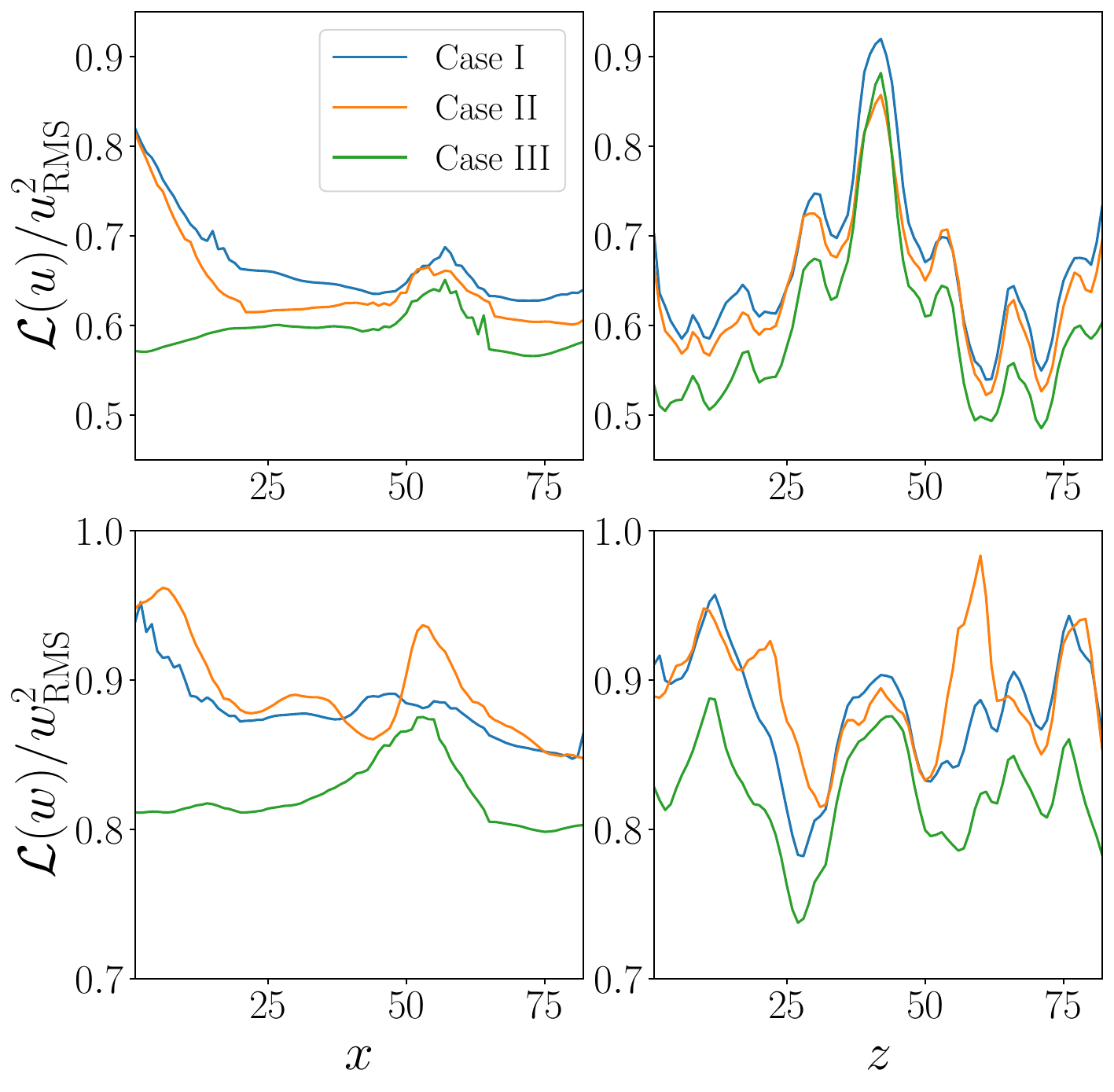}
	\caption{MSE in the prediction of the synthetic DNS fields as a function of the streamwise (left) and spanwise (right) location.}
	\label{fig:padding}
\end{figure}
The region close to the boundaries of the input samples deserve special attention, since the lack of information outside the domain can negatively impact the prediction quality. Since FCNs perform predictions based on local information,~\cite{guastoni_2021} showed that periodic boundary conditions can be enforced analytically by periodically padding the input fields. This approach can improve the results also for other boundary conditions, however, it requires information outside the boundaries of the individual samples. 
Here, we seek to quantify the effect of the boundary treatment on the prediction performance. We consider predictions using the synthetic DNS fields, using three different boundary treatments:
\begin{itemize}
    \item \textbf{Case I:} Small zero-padding in all layers. In this case the filtered fields after each convolution maintain the same size. Input and output have the same size ($82 \times 82$)
    \item \textbf{Case II:} large zero-padding of the input. The filtered fields after each convolution are progressively smaller after the application of a convolution operation. The padding of the initial field depends on the receptive field of the FCN and it is computed to obtain an output field of the desired size. In this case the input has size $120 \times 120$.
    \item \textbf{Case III:} large informative padding of the input. Similar to the previous case, but the padding is performed by considering additional points from the original input fields along the streamwise and spanwise direction. The input size is $120 \times 120$ also in this case.
\end{itemize}
The first two cases provide the same amount of information, in different forms. The resulting error in both the streamwise and spanwise direction follows a similar trend, as shown in figure~\ref{fig:padding}. The additional information at the boundaries significantly improves the predictions close to the edges in both homogeneous directions and velocity components.
Table~\ref{tab:padding} reports the average error on the training dataset. Case I and II provide very similar error values, highlighting how the two procedures are essentially equivalent. By contrast, case III reduces the MSE by about 10\% with respect to Case I and II, allowing for a better reconstruction if the additional information is available.
It should also be noted that the first two approaches yield similar results also after the fine-tuning on the experimental fields.

\begin{table}
    \caption{Normalized MSE comparison for the wall-parallel velocity components in the predictions of DNS fields using different boundary treatments.}
    \label{tab:padding}
    \begin{tabular}{c c c}
    \toprule
    & $\mathcal{L}(u)/u^2_\mathrm{RMS}$ & $\mathcal{L}(w)/w^2_\mathrm{RMS}$ \\[3pt]
    \midrule
    Case I   & 0.667 & 0.880 \\[3pt]
    Case II  & 0.643 & 0.892 \\[3pt]
    Case III & 0.587 & 0.790 \\
    \botrule
    \end{tabular}
\end{table}

\end{appendices}


\bibliography{jfm}

\begin{thebibliography}{39}
\providecommand{\natexlab}[1]{#1}
\providecommand{\url}[1]{{#1}}
\providecommand{\urlprefix}{URL }
\providecommand{\doi}[1]{\url{https://doi.org/#1}}
\providecommand{\eprint}[2][]{\url{#2}}
 \bibcommenthead

\bibitem[{Abe et~al(2004)Abe, Kawamura, and Matsuo}]{ABE2004404}
Abe H, Kawamura H, Matsuo Y (2004) Surface heat-flux fluctuations in a turbulent channel flow up to ${R}e_{\tau} = 1020$ with ${P}r=0.025$ and 0.71. International Journal of Heat and Fluid Flow 25(3):404--419. \doi{https://doi.org/10.1016/j.ijheatfluidflow.2004.02.010}, \urlprefix\url{https://www.sciencedirect.com/science/article/pii/S0142727X04000153}, turbulence and Shear Flow Phenomena (TSFP-3)

\bibitem[{Alc{\'a}ntara-{\'A}vila et~al(2021)Alc{\'a}ntara-{\'A}vila, Hoyas, and P{\'e}rez-Quiles}]{alcantra_0p71}
Alc{\'a}ntara-{\'A}vila F, Hoyas S, P{\'e}rez-Quiles MJ (2021) Direct numerical simulation of thermal channel flow for ${R}e_{\tau} = 5000$ and ${P}r=0.71$. J Fluid Mech 916

\bibitem[{Astarita(2007)}]{Astarita2007}
Astarita T (2007) Analysis of weighting windows for image deformation methods in {PIV}. Exp Fluids 43(6):859--872

\bibitem[{Astarita and Cardone(2004)}]{Astarita2004}
Astarita T, Cardone G (2004) Analysis of interpolation schemes for image deformation methods in {PIV}. Exp Fluids 38(2):233--243

\bibitem[{Astarita and Carlomagno(2012)}]{astarita2012infrared}
Astarita T, Carlomagno GM (2012) Infrared thermography for thermo-fluid-dynamics. Springer Science \& Business Media

\bibitem[{Balasubramanian et~al(2023)Balasubramanian, Guastoni, Schlatter, and Vinuesa}]{Balasubramanian_Guastoni_Schlatter_Vinuesa_2023}
Balasubramanian AG, Guastoni L, Schlatter P, et~al (2023) Direct numerical simulation of a zero-pressure-gradient turbulent boundary layer with passive scalars up to prandtl number pr=6. Journal of Fluid Mechanics 974:A49. \doi{10.1017/jfm.2023.803}

\bibitem[{Bor{\'e}e(2003)}]{boree2003extended}
Bor{\'e}e J (2003) Extended proper orthogonal decomposition: a tool to analyse correlated events in turbulent flows. Exp Fluids 35(2):188--192

\bibitem[{Chevalier et~al(2007)Chevalier, Schlatter, Lundbladh, and Henningson}]{simson}
Chevalier M, Schlatter P, Lundbladh A, et~al (2007) {SIMSON} : a pseudo-spectral solver for incompressible boundary layer flows. Tech. rep., KTH Royal Institute of Technology, Stockholm

\bibitem[{Dumoulin and Visin(2016)}]{dumoulin2016guide}
Dumoulin V, Visin F (2016) A guide to convolution arithmetic for deep learning. {P}reprint arXiv:160307285

\bibitem[{Encinar and Jim\'enez(2019)}]{miguel}
Encinar MP, Jim\'enez J (2019) Logarithmic-layer turbulence: A view from the wall. Phys Rev Fluids 4:114603. \doi{10.1103/PhysRevFluids.4.114603}, \urlprefix\url{https://link.aps.org/doi/10.1103/PhysRevFluids.4.114603}

\bibitem[{Foroozan et~al(2023)Foroozan, Guemes, Raiola, Castellanos, Discetti, and Ianiro}]{foroozan}
Foroozan F, Guemes A, Raiola M, et~al (2023) Synchronized measurement of instantaneous convective heat ﬂux and velocity ﬁelds in wall-bounded ﬂows. Meas Sci Technol [in press] 34(12):125301

\bibitem[{Gautier et~al(2015)Gautier, Aider, Duriez, Noack, Segond, and Abel}]{gautier_2015}
Gautier N, Aider JL, Duriez T, et~al (2015) Closed-loop separation control using machine learning. J Fluid Mech 770:442–457. \doi{10.1017/jfm.2015.95}

\bibitem[{Guastoni et~al(2020)Guastoni, Encinar, Schlatter, Azizpour, and Vinuesa}]{guastoni2020prediction}
Guastoni L, Encinar MP, Schlatter P, et~al (2020) Prediction of wall-bounded turbulence from wall quantities using convolutional neural networks. J Phys: Conf Ser 1522(1):012022

\bibitem[{Guastoni et~al(2021)Guastoni, G\"uemes, Ianiro, Discetti, Schlatter, Azizpour, and Vinuesa}]{guastoni_2021}
Guastoni L, G\"uemes A, Ianiro A, et~al (2021) Convolutional-network models to predict wall-bounded turbulence from wall quantities. J Fluid Mech 928:A27. \doi{10.1017/jfm.2021.812}

\bibitem[{G{\"u}emes et~al(2019)G{\"u}emes, Discetti, and Ianiro}]{guemes2019sensing}
G{\"u}emes A, Discetti S, Ianiro A (2019) Sensing the turbulent large-scale motions with their wall signature. Phys Fluids 31(12)

\bibitem[{G\"uemes et~al(2021)G\"uemes, Discetti, Ianiro, Sirmacek, Azizpour, and Vinuesa}]{guemes_2021}
G\"uemes A, Discetti S, Ianiro A, et~al (2021) From coarse wall measurements to turbulent velocity fields through deep learning. Phys Fluids 33(7):075121. \doi{10.1063/5.0058346}

\bibitem[{Gurka et~al(2004)Gurka, Liberzon, and Hetsroni}]{gurka}
Gurka R, Liberzon A, Hetsroni G (2004) Detecting coherent patterns in a flume by using piv and ir imaging techniques. Exp Fluids 37(2):230--236. \doi{10.1007/s00348-004-0805-3}

\bibitem[{He et~al(2016)He, Zhang, Ren, and Sun}]{resnet}
He K, Zhang X, Ren S, et~al (2016) Deep residual learning for image recognition. In: 2016 IEEE Conf. Comp. Vision and Pattern Recogn. (CVPR), pp 770--778, \doi{10.1109/CVPR.2016.90}

\bibitem[{Hetsroni and Rozenblit(1994)}]{hetsroni1994heat}
Hetsroni G, Rozenblit R (1994) Heat transfer to a liquid—solid mixture in a flume. Int J Multiph Flow 20(4):671--689

\bibitem[{Hinton et~al(2012)Hinton, Srivastava, Krizhevsky, Sutskever, and Salakhutdinov}]{dropout}
Hinton GE, Srivastava N, Krizhevsky A, et~al (2012) Improving neural networks by preventing co-adaptation of feature detectors. CoRR abs/1207.0580. \urlprefix\url{http://arxiv.org/abs/1207.0580}, {\href{https://arxiv.org/abs/1207.0580}{{1207.0580}}}

\bibitem[{Ho et~al(2020)Ho, Jain, and Abbeel}]{diffusion}
Ho J, Jain A, Abbeel P (2020) Denoising diffusion probabilistic models. \textit{ar{X}iv:2006.11239}

\bibitem[{Ioffe and Szegedy(2015)}]{batchnorm}
Ioffe S, Szegedy C (2015) Batch normalization: Accelerating deep network training by reducing internal covariate shift. CoRR abs/1502.03167. \urlprefix\url{http://arxiv.org/abs/1502.03167}, {\href{https://arxiv.org/abs/1502.03167}{{1502.03167}}}

\bibitem[{Kim and Lee(2020)}]{kim_lee_2020}
Kim J, Lee C (2020) Prediction of turbulent heat transfer using convolutional neural networks. J Fluid Mech 882:A18. \doi{10.1017/jfm.2019.814}

\bibitem[{Kingma and Ba(2015)}]{kingmaba}
Kingma DP, Ba J (2015) Adam: {A} method for stochastic optimization. In: 3rd Int. Conf. on Learning Representations, {ICLR} 2015, San Diego, CA, USA, May 7-9, 2015

\bibitem[{Kozuka et~al(2009)Kozuka, Seki, and Kawamura}]{kozuka}
Kozuka M, Seki Y, Kawamura H (2009) {DNS} of turbulent heat transfer in a channel flow with a high spatial resolution. Int J Heat Fluid Fl 30(3):514--524

\bibitem[{Mendez et~al(2017)Mendez, Raiola, Masullo, Discetti, Ianiro, Theunissen, and Buchlin}]{mendez2017pod}
Mendez MA, Raiola M, Masullo A, et~al (2017) Pod-based background removal for particle image velocimetry. Exp Therm Fluid Sci 80:181--192

\bibitem[{Milano and Koumoutsakos(2002)}]{milano2002neural}
Milano M, Koumoutsakos P (2002) Neural network modeling for near wall turbulent flow. J Comput Phys 182(1):1--26

\bibitem[{Nakamura(2009)}]{nakamura2009frequency}
Nakamura H (2009) Frequency response and spatial resolution of a thin foil for heat transfer measurements using infrared thermography. Int J Heat Mass Transf 52(21-22):5040--5045

\bibitem[{Nakamura and Yamada(2013)}]{nakamura}
Nakamura H, Yamada S (2013) Quantitative evaluation of spatio-temporal heat transfer to a turbulent air flow using a heated thin-foil. Int J Heat Mass Transf 64:892--902. \doi{https://doi.org/10.1016/j.ijheatmasstransfer.2013.05.006}, \urlprefix\url{https://www.sciencedirect.com/science/article/pii/S0017931013003967}

\bibitem[{Pan and Yang(2009)}]{pan2009survey}
Pan SJ, Yang Q (2009) A survey on transfer learning. IEEE Trans Knowl Data Eng  22(10):1345--1359

\bibitem[{Raffel et~al(2018)Raffel, Willert, Scarano, K{\"a}hler, Wereley, and Kompenhans}]{raffel2018particle}
Raffel M, Willert CE, Scarano F, et~al (2018) Particle Image Velocimetry: A Practical Guide. Springer

\bibitem[{Raiola et~al(2017)Raiola, Greco, Contino, Discetti, and Ianiro}]{raiola2017towards}
Raiola M, Greco CS, Contino M, et~al (2017) Towards enabling time-resolved measurements of turbulent convective heat transfer maps with {IR} thermography and a heated thin foil. Int J Heat Mass Trans 108:199--209

\bibitem[{Ronneberger et~al(2015)Ronneberger, Fischer, and Brox}]{unet}
Ronneberger O, Fischer P, Brox T (2015) U-net: Convolutional networks for biomedical image segmentation. Springer, pp 234--241

\bibitem[{Sanmiguel~Vila et~al(2017)Sanmiguel~Vila, {\"O}rl{\"u}, Vinuesa, Schlatter, Ianiro, and Discetti}]{sanmiguel2017adverse}
Sanmiguel~Vila C, {\"O}rl{\"u} R, Vinuesa R, et~al (2017) Adverse-pressure-gradient effects on turbulent boundary layers: statistics and flow-field organization. Flow Turbul Combust 99:589--612

\bibitem[{Sasaki et~al(2019)Sasaki, Vinuesa, Cavalieri, Schlatter, and Henningson}]{sasaki_2019}
Sasaki K, Vinuesa R, Cavalieri AVG, et~al (2019) Transfer functions for flow predictions in wall-bounded turbulence. J Fluid Mech 864:708–745. \doi{10.1017/jfm.2019.27}

\bibitem[{Scarano(2001)}]{scarano2001}
Scarano F (2001) Iterative image deformation methods in {PIV}. Meas Sci Technol 13(1):R1

\bibitem[{Soria(1996)}]{SORIA1996}
Soria J (1996) An investigation of the near wake of a circular cylinder using a video-based digital cross-correlation particle image velocimetry technique. Exp Therm Fluid Sci 12(2):221--233

\bibitem[{Vaswani et~al(2017)Vaswani, Shazeer, Parmar, Uszkoreit, Jones, Gomez, Kaiser, and Polosukhin}]{vaswani}
Vaswani A, Shazeer N, Parmar N, et~al (2017) Attention is all you need. \textit{ar{X}iv:1706.03762}. \doi{10.48550/ARXIV.1706.03762}, \urlprefix\url{https://arxiv.org/abs/1706.03762}

\bibitem[{Willert and Gharib(1991)}]{Willert1991}
Willert CE, Gharib M (1991) Digital particle image velocimetry. Exp Fluids 10(4):181--193

\end{thebibliography}

\end{document}